\documentclass[%
 reprint,
 superscriptaddress,
 amsmath,amssymb,
 aps,prl,
 floatfix, longbibliography
]{revtex4-2}
\usepackage{graphicx}
\usepackage{dcolumn}
\usepackage{bm}

\usepackage{float}

\usepackage[dvipsnames]{xcolor}

\usepackage{hyperref}
\hypersetup{colorlinks = true, 
urlcolor  = BrickRed, 
linkcolor = NavyBlue, 
citecolor = ForestGreen}

\usepackage{bbm}

\newcommand{\kb}{k_{B}}

\begin{document}

\title{Universal Bound on Energy Cost of Bit Reset in Finite Time}

\author{Yi-Zheng Zhen}
\affiliation{Shenzhen Institute for Quantum Science and Engineering and Department~of~Physics,Southern University of Science and Technology, Shenzhen 518055, China}
\affiliation{Hefei National Laboratory for Physical Sciences at Microscale and Department of Modern Physics,
University of Science and Technology of China, Hefei, 230026, China}
\author{Dario Egloff}
 \affiliation{Institute of Theoretical Physics, Technische Universit\"at Dresden, D-01062 Dresden, Germany}
 \affiliation{Max  Planck  Institute  for  the  Physics  of  Complex  Systems, N{\"o}thnitzer  Strasse  38,  01187  Dresden,  Germany}
\author{Kavan Modi}
 \affiliation{School of Physics and Astronomy, Monash University, Clayton, Victoria 3800, Australia}
 \affiliation{Shenzhen Institute for Quantum Science and Engineering and Department~of~Physics,Southern University of Science and Technology, Shenzhen 518055, China}
\author{Oscar Dahlsten}
\email{dahlsten@sustech.edu.cn}
\affiliation{Shenzhen Institute for Quantum Science and Engineering and Department~of~Physics,Southern University of Science and Technology, Shenzhen 518055, China}


\begin{abstract}
We consider how the energy cost of bit reset scales with the time duration of the protocol. Bit reset necessarily takes place in finite time, where there is an extra penalty on top of the quasistatic work cost derived by Landauer. This extra energy is dissipated as heat in the computer, inducing a fundamental limit on the speed of irreversible computers. We formulate a hardware-independent expression for this limit in the framework of stochastic processes. We derive a closed-form lower bound on the work penalty as a function of the time taken for the protocol and bit reset error. It holds for discrete as well as continuous systems, assuming only that the master equation respects detailed balance.
\end{abstract}

\maketitle

Bit reset, the setting of an unknown bit to a fixed value, is an elementary operation in irreversible computation. 
Landauer's principle states that bit reset must cost at least $k_{B}T\ln2$ amount of energy, where $k_{B}$ is Boltzmann's constant and $T$ the temperature of the ambient environment, i.e.\ the computer~\cite{Szilard29, Landauer61, Bennett82}.
The energy cost of binary transistor switching is extrapolated to reach this Landauer limit around 2035~\cite{Frank05}.
This cost is a crucial concern:
in the next ten years digital information processors are expected to consume 1/5 of the world's electricity~\cite{Jones18}.
Moreover, the energy expended is dissipated as heat into the computer and computers are already speed constrained by power dissipation tolerance limits, posing a key problem in continuing Moore's law indefinitely~\cite{LentOPS19}.
Apart from this technological importance, Landauer's principle is also a focal point concerning the role of information in thermodynamics~\cite{delRioARDV11, Aberg11, DahlstenRRV11, EgloffDRV15, Taranto+21}
and has also been extended into the quantum regime, see, e.g.,~\cite{reeb_improved_2014, goold_nonequilibrium_2015, lorenzo_landauers_2015, 
ScandiM19, AbiusoMLS20, timpanaro_landauers_2020, miller_quantum_2020}.

To achieve the $k_{B}T\ln2$ limit, quasistatic protocols are required and often assumed possible~\cite{delRioARDV11, Aberg11, EgloffDRV15}.
However, real-life bit reset~\cite{BerutAPCDL12, Orlov12, Jun14, koskiMPA14, Peterson16, Gaudenzi18, Yan18} takes place in finite time.
This necessitates a possibly dramatically higher work cost than $\kb T\ln 2$: there is a finite-time work penalty to bit reset. 
Understanding this penalty is crucial to the energy efficiency of computers in approaching the Landauer limit of irreversible computing.
Generally, the scaling of power dissipation within a time duration $\tau$ has been argued to be inversely linear in the long-time limit for any stochastic process~\cite{SekimotoS97} and, in the slow-driving case, for classical~\cite{SalamonB83} or quantum systems~\cite{ScandiM19, AbiusoMLS20}.
For bit reset, this scaling also matches theoretical~\cite{AurellMM11, ZulkowskiD14, BoydMC18, ProesmansEB20, ProesmansEB20pre} and experimental results~\cite{BerutAPCDL12, Jun14} on a mesoscopic system immersed in a double-well potential.
Particularly, the optimal reset protocol for the double-well case shows that the additional work (beyond the Landauer limit) scales as $1/\tau$ for cases such as vanishing reset error~\cite{ProesmansEB20} or local equilibrium~\cite{ProesmansEB20pre}. 
Meanwhile, a theoretical model for the reset of a qubit indicates a more benign scaling~\cite{BrowneGDV14}. The tension between these results as well as the multitude of possible hardware implementations of bits, including colloidal systems and fluctuating RC circuits~\cite{Shizume95, AurellGMMM12, DianaBE13, ZulkowskiD14, BrowneGDV14, GavrilovB16, DeshpandeGOJ17}, motivates a search for a universal Landauer's principle that holds in finite time regardless of the physical implementation. This will help us understand how far one can minimize the power dissipation, in principle, by altering the hardware, whether with complementary metal–oxide–semiconductor or alternative  platforms~\cite{RoadMap2017}. %
\begin{figure}[!hbt]
\includegraphics[width=0.9\linewidth]{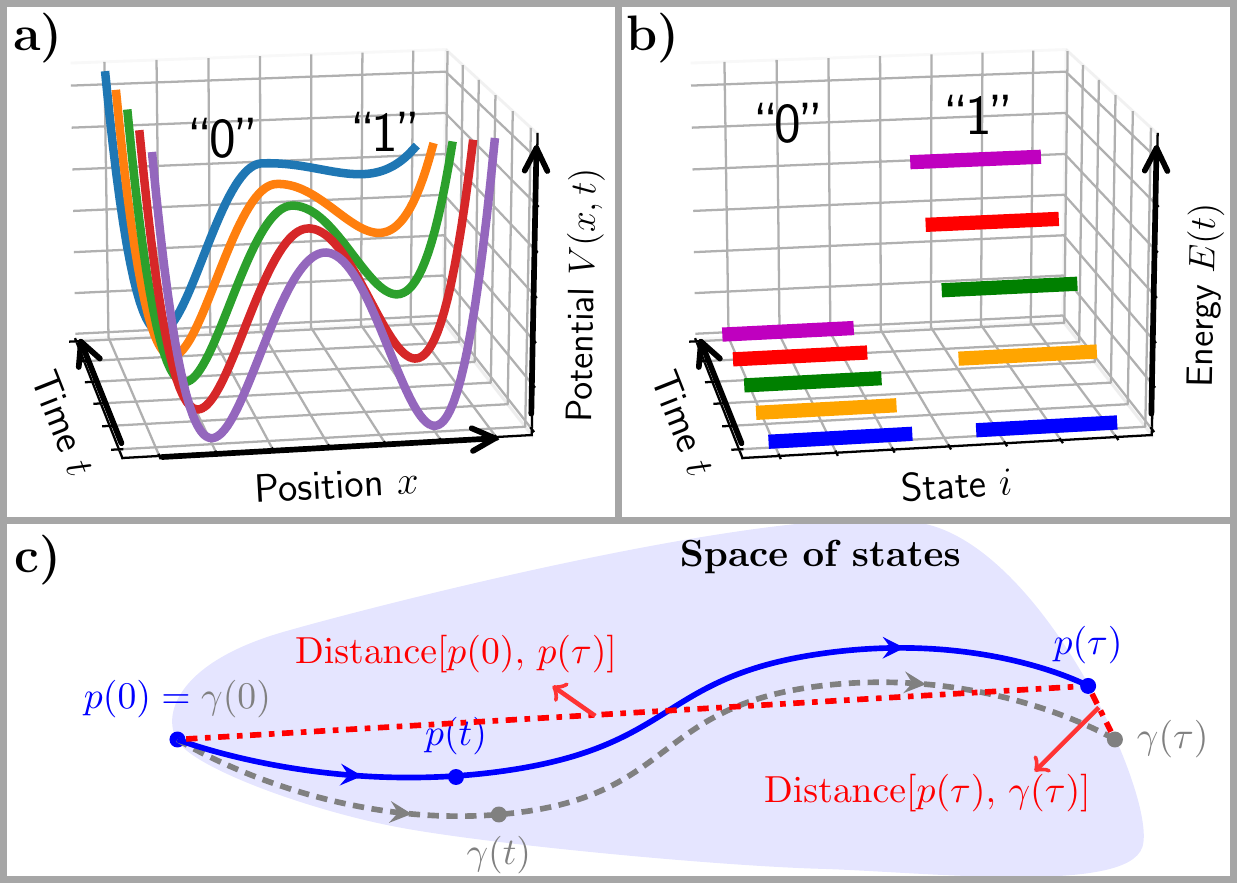}
\caption{\label{fig:schematic}%
Our analysis applies to a wide range of bit-reset scenarios, including: (a) a double-well potential $V(x,t)$ 
which is gradually turned into a single well, (b) a system with two energy levels where
one level is lifted gradually. (c) During the finite-time bit reset, the system state $p(t)$ lags behind the thermal state $\gamma(t)$, and we will show that two distances are closely related to the minimal work cost. }
\end{figure}

In this Letter, we analytically derive a {\em universal} lower bound for the finite-time work penalty in bit reset. The bound applies to all physical models as long as detailed balance is respected and, thus, encompasses previous results focused on specific models such as those depicted in Figs.~\ref{fig:schematic}(a) and~\ref{fig:schematic}(b).
The bound also holds for both the total time region and the finite bit reset error and, thus, can be viewed as a refined Landauer's principle, for finite as well as infinite time.
The proof of the bound implies that the work penalty is optimized if one adopts a truly two-level system, such as a spin.
We further apply the bound to induce an upper limit on the information throughput of irreversible computers.
Our results will aid nanoelectronics research on alternative hardware and in extending information thermodynamics to finite-time nonequilibrium regimes. 

{\em Stochastic thermodynamics.--}%
We briefly introduce common definitions for stochastic thermodynamics using discrete states, though the results straightforwardly apply in the continuous case 
(see~\cite{Gardiner04, Seifert12} and the Supplemental Material~\footnote{See Supplemental Material at [URL] for the precise proof of the main result and for the technical details of the constant-shifting protocol.}).
Denote the state of a physical system by $i\in\{1,2,\dots,N\}$ with an associated energy $E_i$.
We use $p(t)=[p_1(t),\dots,p_N(t)]$ as the distribution of system states at time $t$.
A key assumption is that when the system is in contact with a heat bath at temperature $T$, this distribution undergoes a stochastic evolution toward the 
thermal state $\gamma$, wherein $\gamma_i=e^{-\beta E_i}/Z$ with $\beta=1/k_{B} T$ and the partition function $Z=\sum_ie^{-\beta E_i}$.
The evolution is modeled by a Markovian master equation
\begin{equation}
\label{eq:masterequation}
\frac{d}{dt}p_i\left(t\right) =\sum_{j}\Gamma_{ij}\left(t\right)p_j(t).
\end{equation}
Here, $\Gamma_{ij}(t)$ is the transition rate from state $j$ to state $i$ which satisfies $\Gamma_{ij}(t)\geqslant 0$ for all $ i\neq j$ and $\sum_i\Gamma_{ij}(t)=0$ for any $j$ due to conservation of probability.
We also assume detailed balance
$\Gamma_{ij}(t)\gamma_j(t)=\Gamma_{ji}(t)\gamma_i(t)$
for all $i\neq j$,
which is a common condition satisfied by a wide class of nonequilibrium processes~\cite{Mae20}.

The system is allowed to interact with a work reservoir, which ``drives'' the system~\footnote{The interaction with an ideal work reservoir induces a local time-dependent system Hamiltonian to drive the system.}.
Then, any change in internal energy $U=\sum_i p_i E_i$ is associated with work (due to the work reservoir) and heat (due to the heat bath),
i.e.\ $dU =\sum_i E_i d p_i+\sum_i p_i dE_i=dQ+dW$.
For a process taking time $\tau$, the heat exchanged and work input are then $Q(\tau)=\int_0^{\tau}\dot{Q}dt$ and $W(\tau)=\int_0^{\tau}\dot{W}dt$, respectively.
When the driving is very slow, the system remains in the instantaneous thermal state at all times. Such a process is called quasistatic with associated work $W_{\rm qs}(\tau)$~\footnote{$W_{\rm qs} = \sum_j \int_{E_j(t=0)}^{E_j(t=\tau)} \gamma_j(E_j)dE_j=k_{\rm B}T\ln\frac{Z(0)}{Z(\tau)}$.}.
The $W_{\rm qs}$ can be determined or approximated experimentally either by slowing down the protocol until the work cost is almost constant under further slowing, or in principle, from fast experiments by means of the Jarzynski equality (which follows from detailed balance)~\cite{jarzynski_nonequilibrium_1997}.
The difference between the actual work cost and $W_{\mathrm{qs}}$ can be termed the work penalty
\begin{equation}
\label{eq:work-penalty-def}
W_{\rm pn}\left(\tau\right) = W\left(\tau\right) - W_{\rm qs}(\tau).
\end{equation}
The quasistatic entropy change $\Delta S(\tau)=S(\tau)-S(0)$, where $S=-k_{ B}\sum_i p_i \ln p_i$, is $Q(\tau)/T$, and any additional entropy change is termed the entropy production~\cite{Gardiner04, Seifert12}: $\Sigma(\tau)=\Delta S\left(\tau\right)-Q(\tau)/T\geqslant0$. It is often convenient to use the (non-negative) entropy production rate
\begin{equation}
\label{eq:entropyprodrate}
\frac{d}{dt}\Sigma =\frac{k_{{ B}}}{2}\sum_{i\neq j}\left(\Gamma_{ij}p_{j}-\Gamma_{ji}p_{i}\right)\ln\frac{\Gamma_{ij}p_{j}}{\Gamma_{ji}p_{i}},
\end{equation}
for the case of detailed balance.

{\em Generic bit reset.--}%
Now, we model the bit reset process with stochastic thermodynamics. The bit is associated with
two logical states
``0'' and ``1''.  The bit is {\em reset} (or `erased') if it is set to ``0'' regardless of the initial state. It is normally assumed that either logical state is initially equally likely. 
As can be seen from Fig.~\ref{fig:schematic}(a), logical states may be associated with sets of underlying microstates, ${\Omega}_0$ and ${\Omega}_1$ such that  
\begin{equation}
\label{eq:coarse-grained-state}
P_a^{\rm bit}\left(t\right) = \sum_{i\in{\Omega}_a}p_i\left(t\right),\qquad \forall a\in\left\{0,1\right\}.
\end{equation}
Equation~\eqref{eq:coarse-grained-state} defines the {\em coarse-grained}~\cite{Esposito12} bit distribution  $P^{\rm bit}(t)$. The coarse-grained thermal state can be similarly defined as $\gamma^{\rm bit}$ with $\gamma^{\rm bit}_a=\sum_{i\in{\Omega}_a}\gamma_i$ for $a\in\left\{0,1\right\}$.

By employing interactions with a heat bath and with a work reservoir~\footnote{Note that this is just to account for the energy correctly and that this approach is completely equivalent with allowing energy preserving operations on the system or even to allow any operations that keep the thermal state fixed, where either needs to be supplemented with the possibility of introducing work into the system and accounting for it correctly.},
a reset protocol monotonically increases the energy differences between $\{E_i|i\in{\Omega}_0\}$ and $\{E_j|j\in{\Omega}_1\}$, 
while the system undergoes thermalization.
As depicted in Fig.~\ref{fig:schematic}(a) and~\ref{fig:schematic}(b),
$P_1^{\rm bit}$ decreases leading to a final distribution $P^{\rm bit}(\tau)=[1-\epsilon,\epsilon]$, where $\epsilon=P_1^{\rm bit}(\tau)$ is the reset error. 

Landauer's principle states that the minimal work cost for such a protocol (with $\epsilon\rightarrow 0$) is $W_{\rm qs}=\kb T \ln2$. This can be shown to follow from the non-negativity of entropy production [c.f.\ Eq.~\eqref{eq:penaltyequality} below]. Now, we proceed to bound the work penalty for finite time and nonvanishing reset error.

{\em Main result: Bound on work penalty.--}%
For any bit reset protocols within the framework described above, if initially the system is fully thermalized and $P^{\rm bit}(t=0)=[1/2,1/2]$, the work penalty satisfies
\begin{equation}
\label{eq:penaltybound}
\beta W_{\rm pn}(\tau)\geqslant D_{\epsilon}(\tau)
+\frac{(1-2\epsilon)^2}{\left\langle\mu\right\rangle_{\tau}\tau},
\end{equation}
where $D_{\epsilon}(\tau)=D[P^{\rm bit}(\tau)\|\gamma^{\rm bit}(\tau)]$ is the relative entropy between the coarse-grained bit state with coarse-grained thermal state at the final time and $P^{\rm bit}_1(\tau)=\epsilon$. The reset error $\epsilon$ can be demanded as a boundary condition or it can be treated as a parameter that depends on $\langle\mu\rangle_{\tau}$ for allowed maximal energies. $\langle\mu\rangle_{\tau}$ is a measure of how strong the thermalization is, as discussed later around Eq.~\eqref{eq:entprodspeedlimit}. The proof has three steps
(see Sec.~I of the Supplemental Material~\cite{Note1}).

{Step I: Work penalty from relative entropy}: %
For the (fine-grained) physical systems, the work penalty of any finite-time protocol can be expressed as 
\begin{equation}
\label{eq:penaltyequality}
W_{{\rm pn}}\left(\tau\right)= k_{\rm B}T \Delta D\left[p\|\gamma \right] + T \Sigma\left(\tau\right),
\end{equation}
where $\Delta D[p\|\gamma] = D[p(\tau)\|\gamma(\tau)]-D[p(0)\|\gamma(0)]$ is the change of relative entropy $D[p\|\gamma]=\sum_ip_i\ln \left(p_i/\gamma_i\right)$ (see e.g.\ Ref.~\cite{EspositoB11}). This can be understood from the perspective of information geometry~\cite{ShiraishiS19, Nielsen20}.
As $p(t)$ always chases after $\gamma(t)$ due to thermalization [see Fig.~\ref{fig:schematic}(c)], we examine the rate of change in ``distance'' between them
\begin{equation}
\label{eq:relenttwochanges}
\dot{D}\left[p(t)\|\gamma (t)\right]=\dot{p}\partial_p D\left[p\|\gamma\right]+\dot{\gamma}\partial_\gamma D\left[p\|\gamma\right],
\end{equation}
where $\dot{p}\partial_p D[p\|\gamma]=\sum_i \dot{p_i}\partial_{p_i}D[p\|\gamma]$ and $\dot{\gamma}\partial_\gamma D[p\|\gamma]=\sum_i \dot{\gamma_i}\partial_{\gamma_i}D[p\|\gamma]$.
As proved in
Sec.~IA of the Supplemental Material~\cite{Note1},
the first term is identical to entropy production rate $\dot{\Sigma}$ up to a factor $-k_{B}$, while the second term is exactly the time derivative of $\beta W_{\rm pn}$.
By integrating the equation  $\dot{D}[p\|\gamma]=-\dot{\Sigma}/k_{B}+\beta\dot{W}_{\rm pn}$,  Eq.~\eqref{eq:penaltyequality} is obtained.

{Step 2: Coarse-graining relative entropy, entropy production, and dynamics}: %
The underlying state and dynamics induce corresponding coarse-grained states and dynamics, as described above. 
For general bit reset protocols, the system is initially in a thermal state such that $\Delta D[p\|\gamma] = D[p(\tau)\|\gamma(\tau)]$ in Eq.~\eqref{eq:penaltyequality}.
First, we show that $D[p(\tau)\|\gamma(\tau)]$ can be lower bounded by its coarse-grained counterpart (see Sec.~IB of the Supplemental Material~\cite{Note1})
\begin{equation}
\label{eq:relentcoarse}
D\left[p\left(\tau\right)\|\gamma\left(\tau\right)\right]\geqslant D\left[P^{\rm bit}\left(\tau\right)\|\gamma^{\rm bit}\left(\tau\right)\right] \equiv D_{\epsilon}\left(\tau\right).
\end{equation}

To lower bound entropy production, we require the coarse-grained dynamics.
As a hidden-Markov model
, the coarse-grained dynamics can be described by the master equation $\dot{P}_a^{\rm bit}(t) = \sum_a\Gamma^{\rm bit}_{a\overline{a}}(t)P_{\overline{a}}^{\rm bit}(t)$ for $a\in\{0,1\}$, with a transition rate $\Gamma^{\rm bit}_{a\overline{a}}(t)$ that depends on the underlying microstates~\cite{Esposito12}, which, in turn, are uniquely determined by the initial microstate for the given dynamics.
We define a coarse-grained entropy production rate [c.f.\ Eq.~\eqref{eq:entropyprodrate}]
\begin{equation}
\label{eq:entropyprodcoarse}
\frac{d}{dt}\Sigma^{\rm bit}=\frac{\kb}{2}\sum_{a\in\left\{0,1\right\}}\left(\Gamma_{a\overline{a}}^{{\rm bit}}P_{\overline{a}}-\Gamma_{\overline{a}a}^{{\rm bit}}P_{a}\right)\ln\frac{\Gamma_{a\overline{a}}^{\rm bit}P_{\overline{a}}}{\Gamma_{\overline{a}a}^{\rm bit}P_{a}}.
\end{equation}
Note that $\dot{\Sigma}^{\rm bit}$ is experimentally accessible via measuring $\Gamma_{a\overline{a}}$ following the coarse-grained trajectories, as in, e.g.,~\cite{BerutAPCDL12, koskiMPA14, koski+13, Maillet+2019}.
We prove in Sec.~IB of the Supplemental Material~\cite{Note1} that this $\Sigma^{\rm bit}(\tau)$ is non-negative and is a lower bound to the exact $\Sigma(\tau)$ of the system, i.e.,
\begin{equation}
\label{eq:entprodentprodcg}
\Sigma\left(\tau\right)\geqslant\Sigma^{\rm bit}\left(\tau\right).
\end{equation}
Then, combining Eqs.~\eqref{eq:penaltyequality}, ~\eqref{eq:relentcoarse}, and ~\eqref{eq:entprodentprodcg}, the work penalty of bit reset is always lower bounded by
\begin{equation}
\label{eq:penalty2bit}
W_{\rm pn}(\tau)\geqslant \kb TD\left[P^{\rm bit}\left(\tau\right)\|\gamma^{\rm bit}\left(\tau\right)\right]+T\Sigma^{\rm bit}\left(\tau\right).
\end{equation}

The bound in Eq.~\eqref{eq:penalty2bit} is tight in the case of local equilibrium~\cite{Esposito12, ProesmansEB20pre}, wherein the probabilities of microstates conditional on the bit value are proportional to those of a thermal state (see Sec.~IC of the Supplemental Material~\cite{Note1}).
This also strongly suggests that two-level systems outperform mesoscopic systems in terms of work penalty in general: for any coarse-grained master equation, there exists a qubit evolution such that the work penalty lower bound is saturated.
We have not allowed for the qubit to have quantum coherence here; there are reasons to think it does not help in bit reset protocols~\cite{BrowneGDV14}.

{Step 3: Entropy production time scaling via speed-limit}: %
First, we turn our attention to the $\tau$ dependence of the second term of Eq.~\eqref{eq:penalty2bit} by extending the recently discovered classical  speed limit~\cite{ShiraishiFS18} approach to the coarse-grained case. Speed limits were first proposed to operationalize the time-energy uncertainty relation~\cite{Campaioli2020} and, since then, have found utility in many facets of quantum and classical physics~\cite{
Carlini2006,
Caneva2009,
GiovannettiLM11,
delCampoEPH13,
Brody2015,
Wang2015,
DeffnerC17, 
Campaioli+17, 
OkuyamaO18,
ShanahanCMdC18,
BuscemiFMR20,
Deffner20,
VoVH20}. In essence, the changes to the state of a system must occur in finite time, which depends on the distance between the initial and the final states and the speed, i.e., the system's energetics~\cite{Campaioli2019}.

We show that the coarse-grained master equation can always be mapped to a partial swap model $\dot{P}^{\rm bit}(t)=\mu(t)[P^{\rm bit}(t)-P^{\rm st}(t)]$, with a parameter $\mu(t)=\sum_a\Gamma_{a\overline{a}}^{\rm bit}(t)$ describing the swap rate during relaxation and $P^{\rm st}(t)$ the coarse-grained stationary state~\footnote{The stationary state $P^{\rm st}(t)$ depends on the underlying dynamics and it approaches $\gamma^{\rm bit}(t)$ if the system thermalizes. The partial swap process can be understood via an infinitesimal time $dt$, where the coarse-grained state $P^{\rm bit}(t)$ evolves to $P^{\rm bit}(t+dt)=(1-\mu(t)dt)P^{\rm bit}(t)+\mu(t)dtP^{\rm st}(t)$. This justifies $\mu(t)$ as the rate of swap.} 
(see Sec.~ID of the Supplemental Material~\cite{Note1}).
Using this model, we prove the speed limit for the coarse-grained system,
\begin{equation}
\label{eq:entprodspeedlimit}
\tau \geqslant k_{{\rm B}}\frac{L[P^{\rm bit}(\tau),P^{\rm bit}(0)]^{2}}{\left\langle\mu\right\rangle_{\tau}\Sigma^{\rm bit}(\tau)},
\end{equation}
where $\langle\mu\rangle_{\tau}=\tau^{-1}\int_0^{\tau}dt\mu(t)$ is the time averaged swap rate, and  $L[P^{\rm bit}(\tau),P^{\rm bit}(0)] =\sum_{a}\left|P_{a}^{\rm bit}(\tau)-P_{a}^{\rm bit}(0)\right|$ is the 1-norm distance (see proof in Sec.~IE of the Supplemental Material~\cite{Note1}).
The ``speed'' here is, thus, $\langle\mu\rangle_{\tau}\Sigma^{\rm bit}(\tau)$.
As was the case for $\dot{\Sigma}^{\rm bit}$, $\langle\mu\rangle_{\tau}$ can also be obtained by following the coarse-grained trajectories in experiments via $\mu(t)=\Gamma_{01}^{\rm bit}(t)+\Gamma_{10}^{\rm bit}(t)$. 
Moreover, in Sec.~IF of the Supplemental Material~\cite{Note1} we describe how $\langle\mu\rangle_{\tau}$ can be derived exactly or upper bounded given a master equation.
We remark that the bound is tight in the quasistatic limit, as one can see, by inspection, that both sides of Eq.~\eqref{eq:entprodspeedlimit} go to $0$ in that limit. It also performs well in finite times in the example investigated in Fig.~\ref{fig:BDGVexample} below.

Finally, using Eqs.~\eqref{eq:penalty2bit} and ~\eqref{eq:entprodspeedlimit} 
while substituting the initial condition $P^{\rm bit}(0)=[1/2,1/2]$ and final condition $P^{\rm bit}(\tau)=[1-\epsilon,\epsilon]$ for a generic bit reset protocol, the penalty bound Eq.~\eqref{eq:penaltybound} is deduced.

{\em Relative entropy time scaling.--}%
There are reasons to think that the relative entropy term will, in general, decay exponentially in $\tau$. For example, we derive a general lower bound (in Sec.~II of the Supplemental Material~\cite{Note1}) which has exponential decay for the case of local equilibrium as well as more generally for short times. For the latter case it takes the form 
$D_{\epsilon}(\tau)\geqslant e^{-\langle\mu\rangle_{\tau}\tau}(2e^{-\langle\mu\rangle_{\tau}\tau} - 1) D[\gamma^{\rm bit}(0)\|\gamma^{\rm bit}(\tau)]$.
This is in contrast to the speed limit lower bound $\Sigma^{\rm bit}(\tau)$, which has inverse linear scaling in $\tau$. To investigate this in more depth, now, we specialise to a  typical reset protocol in a two-level system.

{\em Discrete level-shifting protocol in a two-level system.--}%
As we have shown in Eq.~\eqref{eq:penalty2bit}, a two-level system may optimize the work penalty. 
For such a system, e.g., Fig.~\ref{fig:schematic}(b), coarse-grained and fine-grained states coincide,
such that the evolution of the system is described by the partial swap model $\dot{P}(t)=\mu[\gamma(t)-P(t)]$,
where we have assumed $\mu(t)=\mu$ for simplicity. This model is a simplified collisional model for thermalization~\cite{ScaraniZSGB02, BrowneGDV14}, which has been used to
describe relaxation of nuclear magnetic resonance qubits.

A typical reset protocol is to shift one energy level at a constant rate, as discussed in~\cite{BrowneGDV14}.
The two energy levels are both zeros initially, and then the second energy level is shifted to a maximal energy $E_{\max}$ in $N$ steps: for each step,
$E_1$ is suddenly lifted with an increment $E_{\max}/N$, and then, the system thermalizes for a time $\tau/N$. Physically, this assumes the experimentalist can change the control fields significantly faster than the thermalization rate.
We prove that $D_{\epsilon}(\tau)$ in Eq.~\eqref{eq:penaltybound} must have an exponential scaling
(see Sec.~IIIB of the Supplemental Material~\cite{Note1} for the proof),
\begin{equation}
\label{eq:Dupperbound}
D_{\epsilon}\left(\tau\right)\leqslant e^{-\mu \tau /N}D\left[\gamma(0)\|\gamma(\tau)\right].
\end{equation}

\begin{figure}
\centering
\includegraphics[width=\columnwidth]{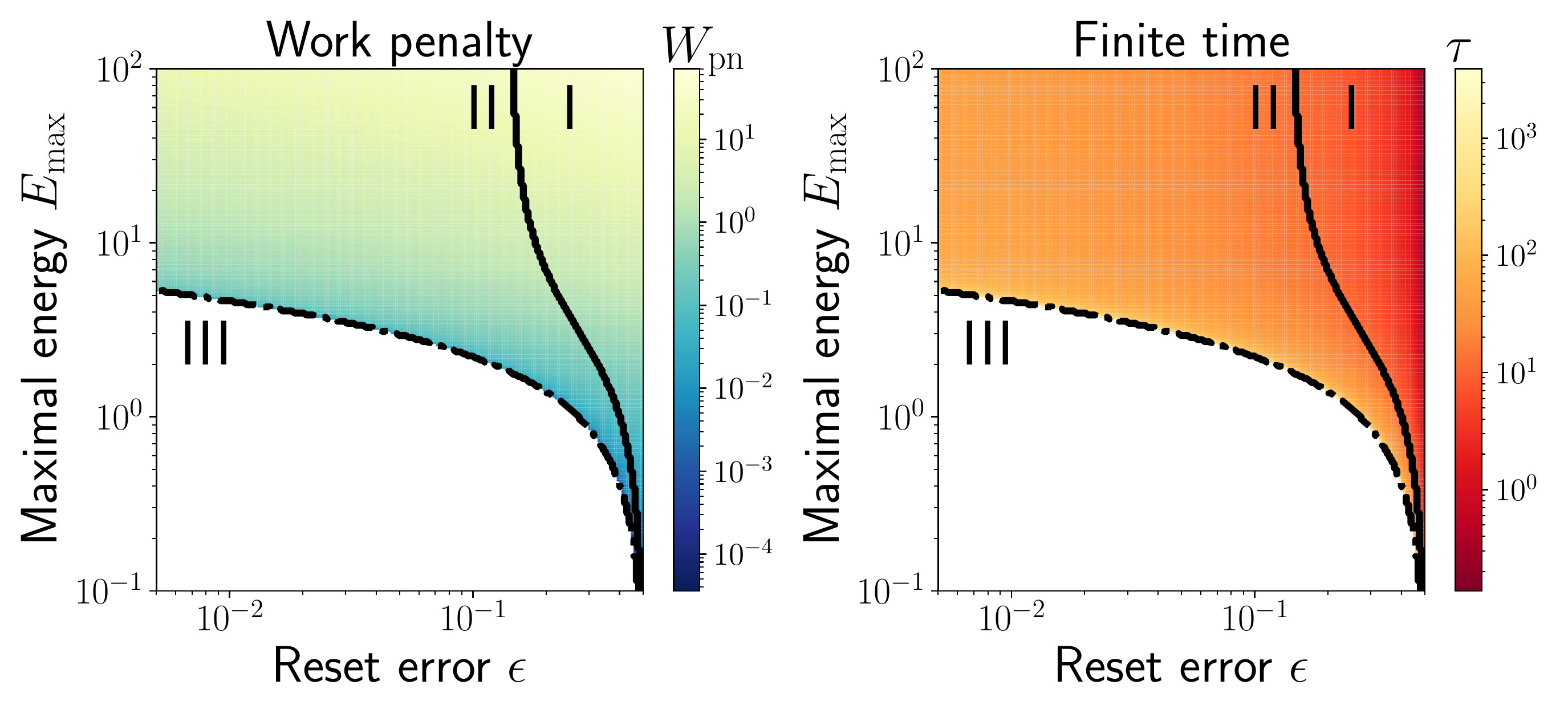}
\caption{\label{fig:BDGVexampleheat}
Domination of $D_{\epsilon}(\tau)$ or $\Sigma(\tau)$ for the work penalty.
Region I: $D_{\epsilon}(\tau)>\Sigma(\tau)$. 
Region II: $D_{\epsilon}(\tau)<\Sigma(\tau)$. 
Region III: the protocol fails to reset the bit given the error and energy. 
Solid line: $D_{\epsilon}(\tau)=\Sigma(\tau)$.
Dash-dotted line:  $\epsilon=\gamma_1(E_{\max})$.
In the simulation, $N=100$, $\mu=0.1$, and $\beta=1$.
}
\end{figure}

Figure~\ref{fig:BDGVexample} gives numerical simulations showing the validity of Eq.~\eqref{eq:penaltybound} and its tightness. 
Two qualitatively different cases are examined: that of fixed maximal energy level $E_{\max}$, and that of fixed reset error $\epsilon$. 
For both cases, the bound in Eq.~\eqref{eq:penaltybound} restricts the work penalty closely and the bound becomes tighter as the finite time gets longer.
The relative entropy and entropy production are upper and lower bounded in Eqs.~\eqref{eq:Dupperbound} and~\eqref{eq:entprodspeedlimit}, respectively. 
We can observe two time regimes:
for short times $D_{\epsilon}(\tau)$ dominates, and $W_{\rm pn}(\tau)$ decays exponentially,
and for long times, $\Sigma(\tau)$ dominates such that $W_{\rm pn}(\tau)$ cannot escape an inverse linear scaling.
As we prove in Sec.~IIID of the Supplemental Material~\cite{Note1},
the work penalty itself can
be bounded by an exponentially decaying function which
converges to a positive value. Therefore, an exponential decay of the work penalty (in $\tau$) can, in certain regimes, dominate the known inverse linear scaling.
\begin{figure}
\centering
\includegraphics[width=.9\columnwidth]{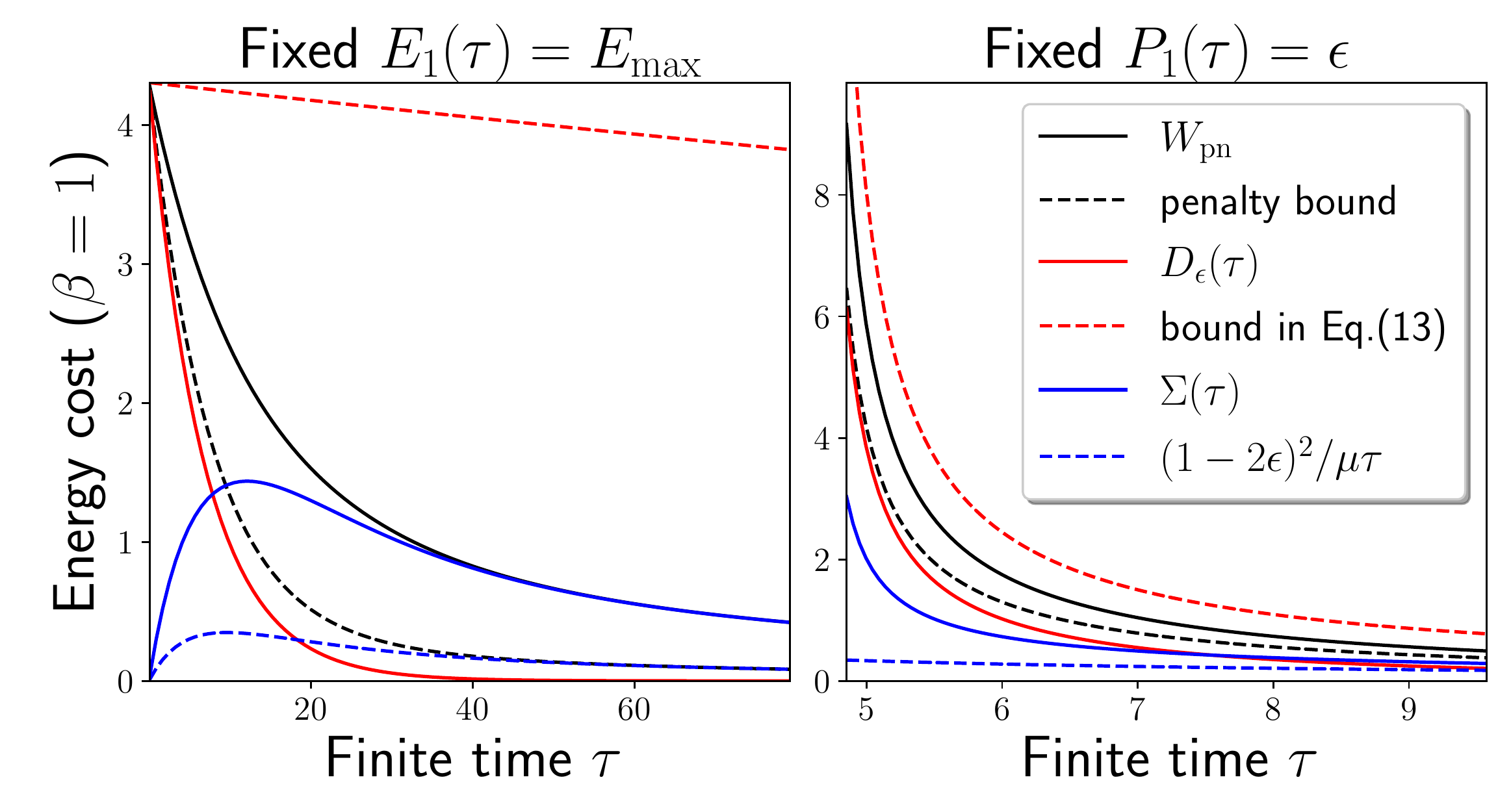}
\caption{\label{fig:BDGVexample}
Time scaling of work penalty, relative entropy, and entropy production for the discrete level-shifting protocol. In the simulation, $N=100$, $\mu=0.1$, $\beta=1$, $E_{\max}=10$ (left), and $\epsilon=0.25$ (right).
}
\end{figure}
To understand the boundaries between the regions where $D_{\epsilon}(\tau)$ or   $\Sigma(\tau)$ dominate the work penalty, in Fig.~\ref{fig:BDGVexampleheat}, we show the performance of the protocol in the entire region of $E_{\max}$ and $\epsilon$. Except for the region where the protocol fails (region III), i.e., where $\epsilon<\gamma_1(E_{\max})$,
the work penalty (finite time) increases (decreases) as both $E_{\max}$ and $\epsilon$ grow.
Roughly speaking, when $\epsilon$ is large, one can always set $E_{\max}$ such that $W_{\rm pn}$ is in an exponential scaling (Region I). When $\epsilon$ is small, $W_{\rm pn}$ is mainly in the inverse linear scaling (Region II).
%

{\em Encompassing previous results.--}%
Our results encompass the overdamped Langevin dynamics treatment of a mesoscopic system in a double-well potential, which is a typical model for bit reset [Fig.~\ref{fig:schematic}(a)]. It describes a large class of mesoscopic systems such as a driven particle that also receives random forces from the environment~\cite{Landauer61, Bennett82, AurellMM11, ZulkowskiD14, ProesmansEB20, ProesmansEB20pre}.
Since the underlying Langevin process is Markovian and respects detailed balance, our bound also applies.

In particular, our results can recover the time scaling demonstrated in important works~\cite{BerutAPCDL12, Jun14, ProesmansEB20, ProesmansEB20pre}.
For experimental works, our results explain the initially observed exponential and, finally, inversely linear work decay~\cite{BerutAPCDL12}, and the entire inversely linear scaling for the perfect reset~\cite{Jun14}.
For theoretical works, our penalty bound  is consistent with recent results based on optimal protocol.
If the reset error  $\epsilon\rightarrow0$, 
$D_{\epsilon}(\tau)$ vanishes, 
the penalty bound 
is reduced to $W_{\rm pn}(\tau)\geqslant \kb T/\langle\mu\rangle_{\tau}\tau$, showing a $1/\tau$ scaling~\cite{ProesmansEB20pre}.
If local equilibrium is assumed, the speed limit bound from Eq.~\eqref{eq:entprodspeedlimit} also shows a $1/\tau$ scaling~\cite{ProesmansEB20}.
For generic cases where previous methods fail to give a closed form of work cost, e.g., with a definite reset error, with a specific landscape of potential, or going beyond the Langevin dynamics, Eq.~\eqref{eq:penaltybound} not only applies to give a refined bound with a $1/\tau$ scaling, but also simplifies the derivation of the bound to the question of finding $\langle \mu\rangle_{\tau}$, which can either be experimentally measured via testing trajectories, or analytically calculated as shown in 
Sec.~IV of the Supplemental Material~\cite{Note1}.

{\em Implied bound on information throughput.--}%
As hardware can handle only limited temperatures
and the work applied for each reset is dissipated as heat~\cite{Frank05}, the power dissipation places a limit on how many resets can be done per unit of time and space.
Using the notation of~\cite{ZhirnovCHB03}, for a device switching time $\tau_{\rm SW}$, integration density (number of binary switches per $\rm cm^2$) $n$, the information throughput $B=n/\tau_{\rm SW}$ is restricted by the ratio of binary transition energy $E_{\rm bit}$ and power dissipation growth $P=E_{\rm bit}B$.
As our results concern the minimal work cost of finite-time bit reset in generic hardwares,
it leads to a refined restriction on the power dissipation growth, i.e.\
$P\geqslant \kb T n [\ln 2 -H_b(\epsilon) +\beta E_{\max}\epsilon +(1-2\epsilon)^2/(\mu \tau_{\rm SW})]/\tau_{\rm SW}$,
where $H_b(\epsilon)=-\epsilon\ln \epsilon-(1-\epsilon)\ln(1-\epsilon)$ is the binary thermodynamic entropy 
(see Sec.~V of the Supplemental Material~\cite{Note1}).

{\em Summary and outlook.--}%
Landauer's principle implies a limit on the information throughput of irreversible computers as real hardware has finite power dissipation tolerance.
Here, we derive a universal version of the finite-time Landauer's principle that applies to any hardware implementation of the bit.
It has been verified with energetic constraints and finite bit reset error thresholds.
It would be interesting to ask if the finite-time Landauer's principle can be extended beyond the detailed balance or, also, into the single-shot regime, i.e.\ statements about the work cost that hold in every single shot of an experiment.
While we mentioned that there are reasons to think that quantum coherence does not allow one to surpass our bound, this should be probed further, e.g.\ using the set-ups of~\cite{reeb_improved_2014, goold_nonequilibrium_2015, lorenzo_landauers_2015, 
ScandiM19, AbiusoMLS20, timpanaro_landauers_2020, miller_quantum_2020}.

\begin{acknowledgments}
We gratefully acknowledge valuable discussions with Konstantin Beyer, Mike Frank, Andrew Garner, Min Jiang, Yi Li, Eric Lutz, Yingqiu Mao, Xin-Hua Peng, Walter Strunz, Vlatko Vedral, Haidong Yuan, and Qunsong Zeng.
Y.Z.Z. and O.D. acknowledge support from the National Natural Science Foundation of China (Grants No.~12050410246, No.~12005091).
D.E. acknowledges support from the Swiss National Science Foundation (Grant~No.~P2SKP2\_18406).
K.M. acknowledges support from the Australian Academy of Technology and Engineering via the 2018 Australia China Young Scientists Exchange Program and the 2019 Next Step Initiative.
Y.Z.Z. also acknowledges support from the China Postdoctoral Science Foundation (Grant No.~2020M671856).
\end{acknowledgments}

\bibliography{ftlrefs}


\end{document}


\title{Supplemental Material:\\
Universal Bound on Energy Cost of Bit Reset in Finite Time}

\author{Yi-Zheng Zhen}
\affiliation{Shenzhen Institute for Quantum Science and Engineering and Department~of~Physics,Southern University of Science and Technology, Shenzhen 518055, China}
\affiliation{Hefei National Laboratory for Physical Sciences at Microscale and Department of Modern Physics,
University of Science and Technology of China, Hefei, 230026, China}
\author{Dario Egloff}
 \affiliation{Institute of Theoretical Physics, Technische Universit\"at Dresden, D-01062 Dresden, Germany}
 \affiliation{Max  Planck  Institute  for  the  Physics  of  Complex  Systems, N{\"o}thnitzer  Strasse  38,  01187  Dresden,  Germany}
\author{Kavan Modi}
 \affiliation{School of Physics and Astronomy, Monash University, Clayton, Victoria 3800, Australia}
 \affiliation{Shenzhen Institute for Quantum Science and Engineering and Department~of~Physics,Southern University of Science and Technology, Shenzhen 518055, China}
\author{Oscar Dahlsten}
\email{dahlsten@sustech.edu.cn}
\affiliation{Shenzhen Institute for Quantum Science and Engineering and Department~of~Physics,Southern University of Science and Technology, Shenzhen 518055, China}

\maketitle


\section{\label{App:secA}%
Proof of the work penalty lower bound}

\subsection{\label{App:secA1}%
Step I: Geometric formulation of work penalty}

For the relative entropy
\begin{equation}
D\left[p\|\gamma\right] = \sum_i p_i \ln\frac{p_i}{\gamma_i},
\label{eq:app-relent}
\end{equation}
here we show that the two parts in its time derivative 
\begin{equation}
\dot{D}\left[p\|\gamma\right]=\dot{p}\partial_p D\left[p\|\gamma\right]+\dot{\gamma}\partial_\gamma D\left[p\|\gamma\right]
\label{eq:app-relentdecomp}
\end{equation}
correspond to entropy production rate and work penalty rate, respectively.

For the first term,
\begin{align}
\dot{p}\partial_p D\left[p\|\gamma\right] &=
	\sum_i \dot{p_i} \partial_{p_i} D\left[p\|\gamma\right]\\
&= \sum_i \dot{p_i}\left( \ln\frac{p_i}{\gamma_i}+1
\right)\\
&= \sum_i \dot{p_i} \left(\ln p_i + \beta E_i + \ln Z
\right )\\
&= \sum_i \dot{p_i} \ln p_i + \beta \sum_i \dot{p_i} E_i
\\
&= -\kb^{-1}\dot{S} + \beta \dot{Q} 
\\
&= -\kb^{-1}\dot{\Sigma},
\label{eq:app-relentdecomp-1}
\end{align}
where we have used the thermal state expression $\gamma_i=\exp(-\beta  E_i)/Z$, the definition of entropy production $\Sigma = \Delta S - Q/T$, and the fact that $\sum_i \dot{p}_i=0$.

For the second term, recall that $W_{\rm pn} = W-W_{\rm qs}$ with $W(\tau)=\int_0^{\tau}dt \sum_i p_i \dot{E_i}$ and $W_{\rm qs}(\tau)=-k_{B}T\ln Z(\tau)/Z(0)$.
We have
\begin{align}
\dot{W}_{\rm pn} &= \sum_i p_i \dot{E_i} + k_{B}T \frac{d}{dt}\ln Z\\
&= k_{B}T \sum_i p_i \frac{d}{dt}\left(\beta E_i + \ln Z\right)\\
&= - k_{B}T \sum_i p_i \frac{d}{dt}\ln\frac{\euler^{-\beta E_i}}{Z}\\
&= - k_{B}T \sum_i p_i \frac{d}{dt}\ln\gamma_i,
\end{align}
such that the second term in Eq.\eqref{eq:app-relentdecomp} is
\begin{equation}
\dot{\gamma}\partial_\gamma D\left[p\|\gamma\right] = 
	-\sum_i p_i \frac{d}{dt}\ln\gamma_i= \beta\dot{W}_{\rm pn}.
\label{eq:app-relentdecomp-2}
\end{equation}
Substituting Eqs.\eqref{eq:app-relentdecomp-1} and \eqref{eq:app-relentdecomp-2} into Eq.\eqref{eq:app-relentdecomp}, we obtain
\begin{equation}
\dot{D}\left[p\|\gamma\right]=-\kb^{-1}\dot{\Sigma}+ \beta\dot{W}_{\rm pn}.
\end{equation}
The integration over time $\tau$ yields
\begin{equation}
W_{\rm pn}\left(\tau\right) = \kb T\Delta D\left[p\|\gamma\right] + T\Sigma\left(\tau\right),
\end{equation}
with $\Delta D[p\|\gamma] = D[p(\tau)\|\gamma(\tau)] - D[p(0)\|\gamma(0)]$.

\paragraph*{Generalization to continuous case}
Making the physically justified assumptions that both $p$ and $\gamma$ are in Schwartz' space (that is, the probabilities are reasonably well localised) and that $\gamma$ has full support (the energy does not diverge), one directly sees that simply replacing the sum by the corresponding integral will give the same result for the continuous case. Of course there are well-motivated, but strictly speaking unphysical, models that do not follow the above assumptions and more care must be taken in those cases. For instance, if the energy does diverge in some volume in space at some time, then $p$ needs to go to zero there before that happens, since otherwise the model introduces infinite work on the system. See Ref.~\cite{Gardiner04} for details.

\subsection{\label{App:secA2}%
Step 2: Coarse-graining relative entropy, entropy production and dynamics}

We apply the coarse-graining to a generic system to create a bit.
Precisely, if the system is observed in one of states in $\Omega_0 = \{i_1,i_2,\dots,i_n\}$, we say the bit is in logical state 0; otherwise, if the system is in $\Omega_1=\{j_1,j_2,\dots,j_m\}$, we say the bit is in logical state 1.
For the bit to be logical, $\Omega_0$ and $\Omega_1$ must be disjoint and complementary.
Then, the distribution of bit states is denoted by $P^{\rm bit}=[P_0^{\rm bit},P_1^{\rm bit}]$ with
\begin{equation}
P_a = \sum_{i\in\Omega_a} p_i, \quad
\forall a\in\left\{0,1\right\}.
\label{eq:app-coarsegrainstate}
\end{equation}
The master equation of coarse-grained states can be written as~\cite{Esposito12}
\begin{equation}
\label{eq:masterequationcoarse}
\frac{d}{dt}P_a^{\rm bit}\left(t\right) = \Gamma^{\rm bit}_{a\overline{a}}\left(t\right)P_{\overline{a}}^{\rm bit}\left(t\right)-\Gamma^{\rm bit}_{\overline{a}a}\left(t\right)P_a^{\rm bit}\left(t\right),
\end{equation}
where $\overline{a}$ is the negation of bit value $a\in\{0,1\}$, and $\Gamma^{\rm bit}_{a\overline{a}} = \sum_{i\in{\Omega}_a}\sum_{j\in{\Omega}_{\overline{a}}}\Gamma_{ij}p_j/P_{\overline{a}}$ can be shown to be the transition rate between coarse-grained states.

\subsubsection{\label{App:secA2a}%
Relative entropy}
We directly apply the log-sum inequality
\begin{equation}
\label{eq:logsumineq}
\sum_i u_i \ln\frac{u_i}{v_i}\geqslant \left(\sum_i u_i\right)\ln\frac{\sum_i u_i}{\sum_i v_i},\quad \forall u_i,v_i\geqslant0,
\end{equation}
where the equality holds when $u_i/v_i$ is a constant for each $i$.
The relative entropy can be bounded as
\begin{align}
D[p\|\gamma]&=\sum_i p_i \ln\frac{p_i}{\gamma_i}=\sum_a \sum_{i\in\Omega_a} p_i \ln\frac{p_i}{\gamma_i}\\
&\geqslant \sum_a \left(\sum_{i\in\Omega_a} p_i\right) \ln\frac{\sum_{i\in\Omega_a} p_i}{\sum_{i\in\Omega_a} \gamma_i}\\
&=\sum_a P_a^{\rm bit}\ln\frac{P_a^{\rm bit}}{\gamma_a^{\rm bit}}\\
&=D\left[P^{\rm bit}\|\gamma^{\rm bit}\right].
\end{align}
For convenience, we use
\begin{equation}
    D_{\epsilon}(\tau)=D\left[P^{\rm bit}(\tau)\|\gamma^{\rm bit}(\tau)\right]
\end{equation}
to denote the relative entropy between the final state and final thermal state, where $\epsilon=P_1^{\rm bit}(\tau)$.
 
\subsubsection{\label{App:secA2b}%
Entropy production}

From the master equation
\begin{equation}
\dot{p}_i=\sum_{j\left(\neq i\right)}\Gamma_{ij}p_j-\Gamma_{ji}p_i,
\end{equation}
where $\Gamma_{ij}$ are transition rates satisfying $\Gamma_{ij}\geqslant0$ for $i\neq j$ and $\sum_i \Gamma_{ij}=0$ for all $j$,
the entropy production has a rate
\begin{align}
\dot{\Sigma} &= \dot{S} - \frac{\dot{Q}}{T} \\
&= -k_{B} \sum_i \dot{p}_i \left(\ln p_i + \beta E_i\right)\\
&= k_{B} \sum_{i \neq j}\left(\Gamma_{ij}p_j -\Gamma_{ji}p_i\right) \ln \frac{\gamma_i}{p_i}\\
&= \frac{k_{B}}{2}\sum_{i \neq j} \left(\Gamma_{ij}p_j -\Gamma_{ji}p_i\right) \ln \frac{\gamma_ip_j}{p_i\gamma_j}\\
&= \frac{k_{B}}{2}\sum_{i \neq j} \left(\Gamma_{ij}p_j -\Gamma_{ji}p_i\right) \ln \frac{\Gamma_{ij}p_j}{\Gamma_{ji}p_i}.
\label{eq:app-entprodrate}
\end{align}
Here, we have used the thermal state expression $\gamma_i = e^{-\beta E_i}/Z$ in the third line and the detailed balance condition $\Gamma_{ij}\gamma_j=\Gamma_{ji}\gamma_i$ in the last line.

Applying the time derivative on Eq.\eqref{eq:app-coarsegrainstate} gives
\begin{align}
\dot{P}_a^{\rm bit} 
&= \sum_{i\in\Omega_a}\dot{p}_i\\
&= \sum_{i\in\Omega_a}\sum_{j\left(\neq i\right)}
    \left(\Gamma_{ij}p_j-\Gamma_{ji}p_i
    \right)\\
&= \sum_{i\in\Omega_a}\sum_{j\left(\neq i\right)}
    \left(\Gamma_{ij}\frac{p_j}{P_{\overline{a}}^{\rm bit}}P_{\overline{a}}^{\rm bit}-\Gamma_{ji}\frac{p_i}{P_a^{\rm bit}}P_a^{\rm bit}
    \right)\\
&= 
\Gamma^{\rm bit}_{a\overline{a}}P_{\overline{a}}^{\rm bit}-\Gamma^{\rm bit}_{\overline{a}a}P_a^{\rm bit}.
\label{eq:app-mastereqcoarsegrain}
\end{align}
Here,
\begin{equation}
\Gamma^{\rm bit}_{a\overline{a}} = \sum_{i\in\Omega_a}\sum_{j\in\Omega_{\overline{a}}}\Gamma_{ij}\frac{p_j}{P_{\overline{a}}^{\rm bit}}
\end{equation}
are transition rates between coarse-grained states, i.e. bit states 0 and 1.

With the coarse-grained transition rates, we can define the coarse-grained entropy production as a similar form of Eq.\eqref{eq:app-entprodrate}, i.e.
\begin{equation}
\dot{\Sigma}^{\rm bit} = \frac{k_{B}}{2}\sum_a \left(\Gamma^{\rm bit}_{a\overline{a}}P^{\rm bit}_{\overline{a}} -\Gamma^{\rm bit}_{\overline{a}a}P^{\rm bit}_a\right) \ln \frac{\Gamma^{\rm bit}_{a\overline{a}}P^{\rm bit}_{\overline{a}}}{\Gamma^{\rm bit}_{\overline{a}a}P^{\rm bit}_a}.
\label{eq:app-entprodcoarsegrain}
\end{equation}
Now we show that $\Sigma\geqslant\Sigma^{\rm bit}$ by proving $\dot{\Sigma}\geqslant\dot{\Sigma}^{\rm bit}\geqslant 0$.

In Eq.\eqref{eq:app-entprodrate}, we can always drop the terms in the sum with $i,j$ in the same $\Omega_a$. This is due to the fact that $\Gamma_{ij}p_j\geqslant 0$ $\forall i \neq j$ and therefore $(\Gamma_{ij}p_j -\Gamma_{ji}p_i) \ln (\Gamma_{ij}p_j/\Gamma_{ji}p_i)\geqslant 0$.
Then, using the log sum inequality again, we have
\begin{align}
\dot{\Sigma} &\geqslant \frac{k_{B}}{2}\sum_a \sum_{i\in\Omega_a}\sum_{j\in\Omega_{\overline{a}}}
 \left(\Gamma_{ij}p_j -\Gamma_{ji}p_i\right) \ln \frac{\Gamma_{ij}p_j}{\Gamma_{ji}p_i}\\
&\geqslant \frac{k_{B}}{2}\sum_a \left(\sum_{i\in\Omega_a,j\in\Omega_{\overline{a}}} \Gamma_{ij}p_j \ln \frac{\sum_{i\in\Omega_a,j\in\Omega_{\overline{a}}}\Gamma_{ij}p_j}{\sum_{i\in\Omega_a,j\in\Omega_{\overline{a}}}\Gamma_{ji}p_i}\right.\\
& \qquad \qquad
+\left.\sum_{i\in\Omega_a,j\in\Omega_{\overline{a}}} \Gamma_{ji}p_i \ln \frac{\sum_{i\in\Omega_a,j\in\Omega_{\overline{a}}} \Gamma_{ji}p_i}{\sum_{i\in\Omega_a,j\in\Omega_{\overline{a}}} \Gamma_{ij}p_j}\right)\\
&= \frac{k_{B}}{2}\sum_a \left(\Gamma^{\rm bit}_{a\overline{a}}P^{\rm bit}_{\overline{a}} -\Gamma^{\rm bit}_{\overline{a}a}P^{\rm bit}_a\right) \ln \frac{\Gamma^{\rm bit}_{a\overline{a}}P^{\rm bit}_{\overline{a}}}{\Gamma^{\rm bit}_{\overline{a}a}P^{\rm bit}_a}\\
&= \dot{\Sigma}^{\rm bit}.
\end{align}
Note that $\dot{\Sigma}\geqslant\dot{\Sigma}^{\rm bit}\geqslant 0$, as each term in the sum is non-negative due to the relation $(u-v)\log(u/v)\geqslant0$ for $u,v>0$.
This also recovers the second law as it shows that the entropy production and its rate is non-negative.
In the bit reset case, since $\Sigma(0)=\Sigma^{\rm bit}(0)=0$ at the beginning, we finally obtain
\begin{equation}
\Sigma(\tau)\geqslant\Sigma^{\rm bit}(\tau),
\label{eq:app-entprod-entprodcg}
\end{equation}
i.e., the coarse-grained entropy production is lower than the fine-grained one.

\subsubsection{\label{App:secA2c}%
Tightness of the bound}
We have show that the work penalty can be lower bounded by
\begin{equation}
\label{eq:app-pen2bit}
W_{\rm pn}\left(\tau\right)\geqslant \kb TD\left[P^{\rm bit}\left(\tau\right)\|\gamma^{\rm bit}\left(\tau\right)\right] + T\Sigma^{\rm bit}\left(\tau\right).
\end{equation}
As we used log-sum inequality in the proof, the minimal work penalty is obtained by letting $p_i/\gamma_i$ be a constant for all $i$ in the same coarse-grained states, i.e.\
\begin{equation}
\frac{p_i}{\gamma_i}=c_a, \quad \forall i\in \Omega_a.
\end{equation}
As we will show in the next subsection, this equation leads to the assumption of local equilibrium, based on which Eq.~\eqref{eq:app-pen2bit} becomes an inequality.

\subsection{\label{App:secD1}%
Effective two-level system in local equilibrium}

We show that by assuming the local equilibrium, the coarse-grained master equation respects detailed balance, such that the coarse-grained bit is reduced to an effective two-level system.
The local equilibrium assumption can be written as
\begin{equation}
	\frac{p_i}{\gamma_i}=\frac{p_j}{\gamma_j}
\end{equation}
whenever $i$ and $j$ are in the same composite bit. Calling $\Omega_0$ the first bit and $\Omega_1$ the second, this is the case if either $i, j \in \Omega_0$ or $i, j \in \Omega_1$. For this special case of local equilibrium, we have a very simple formula for the thermal state $\gamma^{\rm bit}$ of the coarse grained bits, namely
\begin{equation}
	\gamma^{\rm bit}_a:= \sum_{i\in\Omega_a} \gamma_i.
\end{equation}
 For the thermal state, we need to have that it is stationary under the (undriven) master equation, such that 
\begin{equation}
	\frac{\partial}{\partial t} \gamma^{\rm bit}_0=0.
\end{equation}
As the coarse grained system only has two levels, this is also the only condition (unless the dynamics is trivial).
 Using the very useful property that 
\begin{align}\label{Eq:frac_p_gamma}
\frac{P^{\rm bit}_a}{\gamma^{\rm bit}_a}=\frac{\sum_{i\in\Omega_a} p_i}{\sum_{i\in\Omega_a} \gamma_i} = \frac{\sum_{i\in\Omega_a} \frac{p_k}{\gamma_k} \gamma_i}{\sum_{i\in\Omega_a} \gamma_i}
=\frac{p_k}{\gamma_k},
\end{align}
for any $k\in\Omega_a$,
we get indeed
\begin{align}
\frac{\partial}{\partial t} \gamma^{\rm bit}_0 &= \Gamma^{\rm bit}_{00} \gamma_0^{\rm bit}+ \Gamma^{\rm bit}_{01} \gamma_1^{\rm bit}\nonumber\\
&=\sum_{i\in\Omega_0}\left(
		\sum_{j\in\Omega_0} \Gamma_{ij} \frac{p_j}{\sum_{k\in\Omega_0} p_k} \gamma^{\rm bit}_0\right.\nonumber\\
& \qquad \qquad \left.+\sum_{j\in\Omega_1} \Gamma_{ij} \frac{p_j}{\sum_{k\in\Omega_1} p_k} \gamma^{\rm bit}_1
	\right)\nonumber\\
&=\sum_{i\in\Omega_0}\left(
	\sum_{j\in\Omega_0} \Gamma_{ij}  \gamma_j
	+\sum_{j\in\Omega_1} \Gamma_{ij}\gamma_j
	\right)\nonumber\\
&	=\sum_{i\in\Omega_0} \frac{\partial}{\partial t} \gamma_i =0,
\end{align}
as the $\gamma_i$ are the stationary components of the fine grained thermal state. This also means that the $\gamma^{\rm bit}_a$ satisfy the equilibrium condition (i.e.\ the detailed balance)
\begin{align}
	\gamma^{\rm bit}_0 = \frac{\Gamma^{\rm bit}_{01}}{\Gamma^{\rm bit}_{01}-\Gamma^{\rm bit}_{00}}.
\end{align}

With this in mind we can directly see that the coarse grained entropy production is the same as the fine grained one. We start with the observation that
\begin{align}\label{Eq:frac_ln_p_gamma}
&\ln	\frac{
		\Gamma_{10}^{\rm bit}P_{0}
		}{\Gamma_{01}^{\rm bit}P_{1}
	}=
\ln \frac{
	P_{0} (\Gamma_{10}-\Gamma_{00})
}{\Gamma_{01}^{\rm bit}
}+\ln \frac{\Gamma_{10}^{\rm bit}}{P_{1} (\Gamma_{10}-\Gamma_{00})
}
\\&=\ln	\frac{
	P_{0}
}{\gamma^{\rm bit} _0}-\ln	\frac{
P_{1}
}{\gamma^{\rm bit} _1}
\end{align}
and, from Eq.\eqref{eq:app-mastereqcoarsegrain}, we have that
\begin{align}\label{Eq:Padot}
\dot{P}_a 
&= 
\Gamma^{\rm bit}_{a\overline{a}}P_{\overline{a}}-\Gamma^{\rm bit}_{\overline{a}a}P_a =-\dot{P}_{\overline{a}} .
\end{align}
It follows that
\begin{align}
    \frac{d}{dt} \Sigma^{\rm bit}&=\frac{k_{{\rm B}}}{2}\sum_{a\in\left\{0,1\right\}}\left(\Gamma_{a\overline{a}}^{{\rm bit}}P_{\overline{a}}-\Gamma_{\overline{a}a}^{{\rm bit}}P_{a}\right)\ln\frac{\Gamma_{a\overline{a}}^{\rm bit}P_{\overline{a}}}{\Gamma_{\overline{a}a}^{\rm bit}P_{a}}
    \\
    &=k_{{\rm B}}\sum_{a\in\left\{0,1\right\}} \dot{P}_a \ln \frac{P_{a}
}{\gamma^{\rm bit} _a}
    \\&=k_{{\rm B}}\sum_{a\in\left\{0,1\right\}} \sum_{i\in \Omega_a} \dot{p}_i \ln \frac{p_{i}
}{\gamma_i}
\\&=k_{{\rm B}}\sum_{i} \dot{p}_i \ln \frac{p_{i}
}{\gamma_i}
\\&=\frac{d}{dt} \Sigma,
\end{align}
where the first line is the definition, in the second line we used Eq.\eqref{Eq:frac_ln_p_gamma} and \eqref{Eq:Padot}, in the third line we used linearity of the derivative and Eq.\eqref{Eq:frac_p_gamma}.

As for the relative entropy, let\begin{equation}
\frac{p_{i}}{\gamma_{i}}=c_{a},\quad \forall i\in\Omega_{a}.
\end{equation}
We can compare
\begin{align}
D\left[p\|\gamma\right] & =\sum_{a}\sum_{i\in\Omega_{a}}p_{i}\ln\frac{p_{i}}{\gamma_{i}}=\sum_{a}\sum_{i\in\Omega_{a}}p_{i}\ln c_{a}\\
 & =\sum_{a}\sum_{i\in\Omega_{a}}\gamma_{i}c_{a}\ln c_{a}=\sum_{a}\gamma_{a}^{{\rm bit}}c_{a}\ln c_{a}
\end{align}
with
\begin{align}
D\left[P^{{\rm bit}}\|\gamma^{{\rm bit}}\right] & =\sum_{a}P_{a}^{{\rm bit}}\ln\frac{P_{a}^{{\rm bit}}}{\gamma_{a}^{{\rm bit}}}=\sum_{a}\gamma_{a}^{{\rm bit}}c_{a}\ln c_{a}\\
 & =D\left[p\|\gamma\right].
\end{align}
Therefore, for the system satisfying local equilibrium, we can prove that
\begin{equation}
W_{\rm pn}(\tau) = D_{\epsilon}(\tau) +\Sigma^{\rm bit}(\tau)
\end{equation}
such that the penalty inequality is saturated.

\subsection{\label{App:secA3}%
Partial swap model}
When the system is coarse-grained to a bit system, the dynamics of the bit can always be described by the partial swap model.
To see this, we introduce a coarse-grained stationary state $P^{\rm st}$ such that
\begin{equation}
\Gamma^{\rm bit}_{01}P^{\rm st}_1 = \Gamma^{\rm bit}_{10}P^{\rm st}_0,
\end{equation}
and let
\begin{equation}
\mu = \Gamma^{\rm bit}_{01} + \Gamma^{\rm bit}_{10}.
\end{equation}
Then, we have $\Gamma^{\rm bit}_{01}=\mu P_0^{\rm st}$ and $\Gamma^{\rm bit}_{10}=\mu P_1^{\rm st}$.
Substitute it into Eq.\eqref{eq:app-mastereqcoarsegrain} and we have
\begin{align}
\dot{P}^{\rm bit}_0 &= \mu P_0^{\rm st}P_1^{\rm bit} - \mu P_1^{\rm st}P_0^{\rm bit}= \mu \left( P_0^{\rm st}-P_0^{\rm bit}\right)\\
\dot{P}{\rm bit} &= \mu P_1^{\rm st}P_0^{\rm bit} - \mu P_0^{\rm st}P_1^{\rm bit}= \mu \left( P_1^{\rm st}-P_1^{\rm bit}\right).
\end{align}
Finally, Eq.\eqref{eq:app-mastereqcoarsegrain} can be equivalently written as
\begin{equation}
\dot{P}^{\rm bit}(t) = \mu(t) \left( P^{\rm st}(t)-P^{\rm bit}(t)\right).
\label{eq:app-partialswap}
\end{equation}
To see the meaning of partial swap, consider an infinitesimal time $dt$.
A state $P^{\rm bit}(t)$ evolves to $P^{\rm bit}(t+dt) = (1-\mu(t)dt)P^{\rm bit}(t)+\mu(t)dtP^{\rm st}(t)$.
That is, $P^{\rm bit}(t+dt)$ is obtained by swapping the state $P^{\rm bit}(t)$ with the stationary state $P^{\rm st}$ at a probability $\mu(t)dt$.
Therefore, $\mu(t)$ is called the partial swap rate.

\subsection{\label{App:secA4}%
Step 3: Entropy production time-scaling via speed}
Here, we prove the entropy production lower bound via the speed limit on the partial swap model.

Take Eq.\eqref{eq:app-partialswap} into Eq.\eqref{eq:app-entprodcoarsegrain}, we have the coarse-grained entropy production rate
\begin{align}
\dot{\Sigma}^{\rm bit} &= \mu k_{B}\sum_{a\in\left(0,1\right)}\left(P_a^{\rm bit}-P_a^{\rm st}\right)\ln\frac{P_a^{\rm bit}}{P_a^{\rm st}}\\
&\geqslant 2\mu k_{B}\sum_{a\in\left(0,1\right)}\frac{\left(P_a^{\rm bit}-P_a^{\rm st}\right)^2}{P^{\rm bit}_a+P_a^{\rm st}},
\end{align}
where we have used $(u-v)\ln(u/v)\geqslant 2(u-v)^2/(u+v)$ for any $u,v\geqslant0$.
To obtain the speed limit, consider the norm-1 distance 
\begin{align}
&L\left[P^{\rm bit}\left(\tau\right),P^{\rm bit}\left(0\right)\right]\\
=& \sum_{a}\left|P^{\rm bit}_{a}\left(\tau\right)-P^{\rm bit}_{a}\left(0\right)\right|
= \sum_{a}\left|\int_{0}^{\tau}dt\dot{P}^{\rm bit}_{a}\right|\\
\leqslant&
\int_{0}^{\tau}dt\sum_{a}\left|\dot{P}^{\rm bit}_{a}\right|= \int_{0}^{\tau}dt\sum_{a}\left|\mu\left(P_{a}^{{\rm st}}-P_{a}^{\rm bit}\right)\right|\\
=&\int_{0}^{\tau}dt\sum_{a}\sqrt{\mu^{2}\frac{\left(P_{a}^{{\rm st}}-P_{a}^{\rm bit}\right)^{2}}{P_{a}^{{\rm st}}+P^{\rm bit}_{a}}}\sqrt{P_{a}^{{\rm st}}+P_{a}^{\rm bit}}\\
\leqslant& \int_{0}^{\tau}dt\sqrt{\mu^{2}\sum_{a}\frac{\left(P_{a}^{{\rm st}}-P_{a}^{\rm bit}\right)^{2}}{P_{a}^{{\rm st}}+P^{\rm bit}_{a}}}\sqrt{\sum_{a}\left(P_{a}^{{\rm st}}+P^{\rm bit}_{a}\right)}\\
=& \int_{0}^{\tau}dt\sqrt{\frac{\mu}{k_{{\rm B}}}\dot{\Sigma}^{\rm bit}}\\
\leqslant& \sqrt{\frac{\langle\mu\rangle_{\tau}\tau}{k_{{\rm B}}}\Sigma^{\rm bit}(\tau)}.
\end{align}
Here, we have used the Cauchy-Schwartz inequality twice, and $\langle\mu\rangle_{\tau}=\tau^{-1}\int_0^{\tau} dt\mu(t)$ is the time average of partial swap rate.
The speed limit is therefore expressed as
\begin{equation}
\tau\geqslant\frac{k_{B}L\left[P^{\rm bit}\left(\tau\right),P^{\rm bit}\left(0\right)\right]^2}{\langle\mu\rangle_{\tau}\Sigma^{\rm bit}}.
\end{equation}
Combined with $\Sigma(\tau)\geqslant\Sigma^{\rm bit}(\tau)$, we have
\begin{equation}
\Sigma(\tau)\geqslant\frac{k_{B}L\left[P^{\rm bit}\left(\tau\right),P^{\rm bit}\left(0\right)\right]^2}{\langle\mu\rangle_{\tau}\tau}.
\end{equation}

\subsection{Determining the partial swap rate}

We now show how to determine or bound the coarse-grained) partial swap rate $\mu(t)=\Gamma_{01}^{{\rm bit}}(t)+\Gamma_{10}^{{\rm bit}}(t)$ given the fine-grained master equation. We firstly show how to determine it exactly for two cases of master equations and then give an upper bound for the general case.

\subsubsection{Case of partial swap}

We consider firstly the case where the fine-grained system satisfies the partial swap model, i.e.
\begin{equation}
\dot{p}_{i}(t)=\mu(t)\left(\gamma_{i}(t)-p_{i}(t)\right),\qquad\text{for }i=1,2,\dots,N.
\end{equation}
It can be written as the master equation
\begin{equation}
\dot{p}_{i}(t)=\sum_{j}\Gamma_{ij}(t)p_{j}(t),\qquad\Gamma_{ij}(t)=\mu(t)\left(\gamma_{i}(t)-\delta_{ij}\right).
\end{equation}
Then, for the coarse-grained states
\begin{equation}
P_{a}^{{\rm bit}}(t)=\sum_{i\in\Omega_{a}}p_{i}(t),
\end{equation}
the coarse-grained transition rates can be derived as
\begin{align}
\Gamma_{a\overline{a}}^{{\rm bit}}(t) & =\sum_{i\in\Omega_{a}}\sum_{j\in\Omega_{\overline{a}}}\frac{\Gamma_{ij}(t)p_{j}(t)}{P_{\overline{a}}^{{\rm bit}}(t)}\\
 & =\mu(t)\sum_{i\in\Omega_{a}}\sum_{j\in\Omega_{\overline{a}}}\frac{\left(\gamma_{i}(t)-\delta_{ij}\right)p_{j}(t)}{P_{\overline{a}}^{{\rm bit}}(t)}\\
 & =\mu(t)\sum_{i\in\Omega_{a}}\gamma_{i}\sum_{j\in\Omega_{\overline{a}}}\frac{p_{j}(t)}{P_{\overline{a}}^{{\rm bit}}(t)}\\
 & =\mu(t)\gamma_{a}^{{\rm bit}}(t),\\
\Gamma_{aa}^{{\rm bit}}(t) & =-\Gamma_{\overline{a}a}^{{\rm bit}}(t)\\
 & =\mu(t)\left(\gamma_{a}^{{\rm bit}}(t)-1\right),
\end{align}
for $a\in{0,1}$.
Therefore, the coarse-grained master equation can be transferred to the
partial swap model with the same partial swap rate $\mu$, i.e.\
\begin{equation}
\dot{P}_{a}^{{\rm bit}}(t)=\mu(t)\left(\gamma_{a}^{{\rm bit}}(t)-P_{a}^{{\rm bit}}(t)\right).
\end{equation}
The time averaged partial swap rate is $\langle\mu\rangle_{\tau}=\tau^{-1}\int_0^{\tau}\mu(t)dt$.
\subsubsection{Case of local equilibrium}
In the second case, we consider the local equilibrium where the system state in a coarse-grained set is proportional to the thermal state, i.e. $p_i(t)/\gamma_i(t)=p_j(t)/\gamma_j(t)$ for $i,j\in\Omega_{a}$.
Then, the coarse-grained transition rates can be expressed as
\begin{align}
\Gamma_{a\overline{a}}^{{\rm bit}}(t) &= \sum_{i\in\Omega_a}\sum_{j\in\Omega_{\overline{a}}}\frac{\Gamma_{ij}(t)p_j(t)}{P^{\rm bit}_{\overline{a}}(t)}\\
&= \sum_{i\in\Omega_a}\sum_{j\in\Omega_{\overline{a}}}\frac{\Gamma_{ij}(t)\gamma_j(t)}{\gamma^{\rm bit}_{\overline{a}}(t)}.
\end{align}
The corresponding partial swap rate is then as
\begin{align}
\mu(t)&=\Gamma_{01}^{{\rm bit}}(t)+\Gamma_{10}^{{\rm bit}}(t)\\
&= \sum_{i\in\Omega_0}\sum_{j\in\Omega_1}\frac{\Gamma_{ij}(t)\gamma_j(t)}{\gamma^{\rm bit}_1(t)}+\frac{\Gamma_{ij}(t)\gamma_j(t)}{\gamma^{\rm bit}_0(t)}.
\end{align}
Therefore, $\mu(t)$ can be obtained if knowing the fine-grained thermal state $\gamma_i(t)$ and the transition rates $\Gamma_{ij}(t)$. In this case, one does not need to solve the master equation to obtain the system states $p_i(t)$.

In fact, one can further simplify the derivation since the coarse-grained transition rates satisfy the detailed balance. As we have shown in Sec. I C.,
\begin{equation}
\Gamma_{01}^{{\rm bit}}\gamma_1^{\rm bit}(t) = \Gamma_{10}^{{\rm bit}}\gamma_0^{\rm bit}(t).
\end{equation}
We define a time-dependent parameter $\omega(t)$ describing the thermalization speed in terms of the coarse-grained transition rates by the expression 
\begin{equation}
\omega(t) = \Gamma_{a\overline{a}}^{\rm bit}(t) /\gamma_a^{\rm bit}(t), \qquad\text{for }a=0,1.
\end{equation}
This gives the partial swap rate
\begin{equation}
\mu(t)=\Gamma_{01}^{{\rm bit}}(t)+\Gamma_{10}^{{\rm bit}}(t)=\omega(t).
\end{equation}
Therefore, it is sufficient to obtain $\mu(t)$ by solely looking at $\omega(t)=\Gamma_{01}^{\rm bit}(t)/\gamma_0^{\rm bit}(t)$, and the time averaged partial swap rate is $\langle\mu\rangle_{\tau}=\tau^{-1}\int_0^{\tau}\omega(t)dt$.
\subsubsection{General discrete case}
For the third case, we consider the general discrete system and upper bound $\mu$ by the underlying dynamics, i.e.\
\begin{align}
\mu(t) & =\Gamma_{01}^{{\rm bit}}(t)+\Gamma_{10}^{{\rm bit}}(t)\\
&=\sum_{i\in\Omega_{0}}\sum_{j\in\Omega_{1}}\frac{\Gamma_{ij}(t)p_{j}(t)}{P_{1}^{{\rm bit}}(t)}+\frac{\Gamma_{ji}(t)p_{i}(t)}{P_{0}^{{\rm bit}}(t)}\\
 & \leqslant\sum_{i\in\Omega_{0}}\max_{j\in\Omega_{1}}\Gamma_{ij}(t)+\sum_{j\in\Omega_{1}}\max_{i\in\Omega_{0}}\Gamma_{ji}(t).
\end{align}
In this manner, the time-averaged partial swap rate can be also upper bounded in terms of the fine-grained master equation transition rates.

Finally we remark that for the continuous system, the transition rate $\Gamma_{x,x^{\prime}}$, where $x,x^{\prime}$ are the continuous variable, may diverge so one needs to consider a region around $x$. We will develop this analysis in follow-up work.

\section{\label{App:secB}%
Time-scaling of relative entropy}

As the partial swap model in Eq.\eqref{eq:app-partialswap} is a first-order
partial differential equation, we have a solution
$P^{{\rm bit}}\left(t\right)=\left[P_{0}^{{\rm bit}}\left(t\right),P_{1}^{{\rm bit}}\left(t\right)\right]$
with
\begin{equation}
P_{1}^{{\rm bit}}\left(t\right)={\rm e}^{-\left\langle \mu\right\rangle _{t}t}\left[P_{1}^{{\rm bit}}\left(0\right)+\int_{0}^{t}ds\,{\rm e}^{\left\langle \mu\right\rangle _{s}s}\mu\left(s\right)P_{1}^{{\rm st}}\left(s\right)\right].
\end{equation}
During the bit reset, $P_{1}^{{\rm st}}\left(s\right)\geqslant P_{1}^{{\rm st}}\left(t\right)$
for any two times $s\leqslant t$. Then, $P_{1}^{{\rm bit}}\left(t\right)$
can be lower bounded by
\begin{align}
P_{1}^{{\rm {\rm bit}}}\left(t\right) & \geqslant{\rm e}^{-\left\langle \mu\right\rangle _{t}t}P_{1}^{{\rm {\rm {\rm bit}}}}\left(0\right)+{\rm e}^{-\left\langle \mu\right\rangle _{t}t}P_{1}^{{\rm st}}\left(t\right)\int_{0}^{t}ds\,{\rm e}^{\left\langle \mu\right\rangle _{s}s}\mu\left(s\right)\\
 & ={\rm e}^{-\left\langle \mu\right\rangle _{t}t}P_{1}^{{\rm {\rm {\rm bit}}}}\left(0\right)+\left(1-{\rm e}^{-\left\langle \mu\right\rangle _{t}t}\right)P_{1}^{{\rm st}}\left(t\right),\label{eq:P1taubound}
\end{align}
where $\int_{0}^{t}ds\,{\rm e}^{\left\langle \mu\right\rangle _{s}s}\mu\left(s\right)={\rm e}^{\left\langle \mu\right\rangle _{t}t}-1$
has been used. To bound $D\left[P^{{\rm bit}}\|\gamma^{{\rm bit}}\right]$,
we need to know how $D\left[P^{{\rm bit}}\|\gamma^{{\rm bit}}\right]$
varies with $P_{1}^{{\rm bit}}$. We prove the following relation.

\vspace{1mm}
\noindent{\bf Proposition.}\emph{
For three binary distributions $p=\left[p_{0},p_{1}\right]$, $x=\left[x_{0},x_{1}\right]$
and $y=\left[y_{0},y_{1}\right]$, if $p_{0}\geqslant p_{1}$ and
$p_{1}\geqslant\alpha x_{1}+\left(1-\alpha\right)y_{1}$ holds for
some $\alpha\in\left[0,1\right]$, then, for another binary distribution
$q=\left[q_{0},q_{1}\right]$ with $q_{1}\leqslant p_{1},$the following
inequality holds,
\begin{align}
D\left[p\|q\right]\geqslant & \alpha D\left[x\|q\right]+\left(1-\alpha\right)D\left[y\|q\right]\nonumber \\
 & +\left(\alpha^{2}-\alpha\right)\left(D\left[x\|y\right]+D\left[y\|x\right]\right).\label{eq:relentinequality}
\end{align}
}

\noindent{\bf Proof.} Note that $D\left[p\|q\right]$ is monotonicly
increasing with $p_{1}$. This can be checked by deriving $\partial_{p_{1}}D\left[p\|q\right]$
and considering its boundaries $q_{1}\leqslant p_{1}\leqslant1/2$. Then,
let $\beta=1-\alpha$ and we have 
\begin{align*}
D\left[p\|q\right]\geqslant & D\left[\alpha x+\beta y\|q\right]\\
= & \sum_{k=0,1}\left(\alpha x_{k}+\beta y_{k}\right)\log\frac{\alpha x_{k}+\beta y_{k}}{q_{k}}\\
= & \alpha\sum_{k}x_{k}\log\frac{\alpha x_{k}+\beta y_{k}}{x_{k}}\frac{x_{k}}{q_{k}}\\
 & +\beta\sum_{k}y_{k}\log\frac{\alpha x_{k}+\beta y_{k}}{y_{k}}\frac{y_{k}}{q_{k}}\\
= & -\alpha D\left[x\|\alpha x+\beta y\right]+\alpha D\left[x\|q\right]\\
 & -\beta D\left[y\|\alpha x+\beta y\right]+\beta D\left[y\|q\right]\\
\geqslant & \ \alpha D\left[x\|q\right]+\beta D\left[y\|q\right]\\
 & -\alpha\beta\left(D\left[x\|y\right]+D\left[y\|x\right]\right).
\end{align*}
In the last inequality, we have used the convexity of relative entropy. \emph{[QED.]}

Now, substitute Eq.\eqref{eq:P1taubound} into Eq.\eqref{eq:relentinequality},
we have
\begin{align}
D_{\epsilon}\left(\tau\right)= & D\left[P^{{\rm bit}}\left(\tau\right)\|\gamma^{{\rm bit}}\left(\tau\right)\right]\nonumber \\
\geqslant & q D\left[\gamma^{{\rm {\rm {\rm bit}}}}\left(0\right)\|\gamma^{{\rm bit}}\left(\tau\right)\right]\nonumber \\
& +\left(1-q\right)D\left[P^{{\rm st}}\left(\tau\right)\|\gamma^{{\rm bit}}\left(\tau\right)\right]\nonumber \\
 & -q \left(1-q\right)J\left[\gamma^{\rm bit}\left(0\right), P^{{\rm st}}\left(\tau\right)\right],
\end{align}
where $q={\rm e}^{-\langle \mu\rangle _{\tau}\tau}$ and $J[r,s]=D[r\|s]+D[r\|s]$ is the symmetric relative entropy.
We have the following cases:
\begin{itemize}
\item Case 1: for small enough time $\tau$ such that $1-{\rm e}^{-\left\langle \mu\right\rangle _{\tau}\tau}$
almost vanishes, up to the first order of $\left(1-e^{-\left\langle \mu\right\rangle _{\tau}\tau}\right)$,
we have 
\begin{align}
D_{\epsilon}\left(\tau\right)\geqslant & e^{-\left\langle \mu\right\rangle _{\tau}\tau}D\left[\gamma^{{\rm {\rm {\rm bit}}}}\left(0\right)\|\gamma^{{\rm bit}}\left(\tau\right)\right]\nonumber \\
 & -\left(1-e^{-\left\langle \mu\right\rangle _{\tau}\tau}\right)D\left[\gamma^{{\rm {\rm {\rm bit}}}}\left(0\right)\|P^{{\rm st}}\left(\tau\right)\right]\nonumber \\
 & +{\cal O}\left(\left(1-e^{-\left\langle \mu\right\rangle _{\tau}\tau}\right)^{2}\right).
\end{align}
As in this case $1 \gtrsim e^{-\left\langle \mu\right\rangle _{\tau}\tau}\gg1-{\rm e}^{-\left\langle \mu\right\rangle _{\tau}\tau}$,
$D_{\epsilon}\left(\tau\right)$ is mainly in an exponential scaling.
\item Case 2: if $P^{{\rm st}}\left(\tau\right)=\gamma^{{\rm bit}}\left(\tau\right)$,
we have
\begin{align}
D_{\epsilon}\left(\tau\right)\geqslant & q^2 J\left[\gamma^{{\rm {\rm {\rm bit}}}}\left(0\right),\gamma^{{\rm bit}}\left(\tau\right)\right]\nonumber \\
 & -qD\left[\gamma^{{\rm bit}}\left(\tau\right)\|\gamma^{{\rm {\rm {\rm bit}}}}\left(0\right)\right]\\
\geqslant&
q
\left(2q-1\right)D\left[\gamma^{{\rm bit}}\left(0\right)\|\gamma^{{\rm {\rm {\rm bit}}}}\left(\tau\right)\right]
\end{align}
where we have used $D[r\|s]\geqslant D[s\|r]$ for two binary distribution with $1/2\geqslant r_1\geqslant s_1$.

To show this, one can consider a function $g\left(r_{1}\right)=D\left[r\|s\right]-D\left[s\|r\right]$ and check its derivatives on $r_{1}$. It yields $\partial_{r_{1}}g\geqslant0$
and $\partial_{r_{1}}^{2}g\geqslant0$ for $s_{1}\leqslant r_{1}\leqslant1/2$.
Thus, $g\left(r_{1}\right)\geqslant0$ leads to $D\left[r\|s\right]\geqslant D\left[s\|r\right]$.
\end{itemize}

\section{\label{App:secC}%
Constant-shifting protocol on a two-level system}

As described in the main text, the constant-shifting protocol has $N$ steps.
Initially, $E_0(0)=E_1(0)=0$ and $P_0(0)=P_1(0)=1/2$.
Then, $E_0(t)=0$ is fixed during a finite time $\tau$, while at $t=(k-1)\tau/N$ for each $k=1,2,\dots,N$, the energy $E_1$ is instantly lifted with an increment $E_{\max}/N$.
The system undergoes a thermalization in the following $\Delta \tau=\tau/N$ time.
At the $k$-th step, the energy $E_1$ is denoted by $E^k_1$, the thermal state is denoted by $\gamma^k$, and the system state is denoted by $P^k$.

We have shown that the relative entropy change is equivalent to the sum of work penalty and the negative entropy production. 
Thus, for each energy lift step, entropy production is zero and relative entropy change is exactly work penalty, while for each thermalization step, work penalty is zero and relative entropy change is exactly the negative entropy production.
To summarize, the entire entropy production is
\begin{equation}
\Sigma=\sum_{k=1}^{N}\Sigma^{k}=\sum_{k=1}^{N}D\left[P^{k-1}\|\gamma^{k}\right]-D\left[P^{k}\|\gamma^{k}\right],
\end{equation}
while the entire work penalty is
\begin{align}
W_{{\rm pn}} & =\sum_{k=1}^{N}W_{{\rm pn}}^{k}=\sum_{k=1}^{N}D\left[P^{k-1}\|\gamma^{k}\right]-D\left[P^{k-1}\|\gamma^{k-1}\right]\\
 & =D\left[P^{N}\|\gamma^{N}\right]+\Sigma.
\end{align}

\subsection{\label{App:secC1}%
Relative entropy production bound with a fixed thermal state}
We first look at the partial swap model $\dot{p}=\mu\left(\gamma-p\right)$ with a time-independent $\gamma$.
This corresponds to each thermalsation step.
The $p(t)$ can be solved as
\begin{equation}
p\left(t\right)=e^{-\mu t}p\left(0\right)+\left(1-e^{-\mu t}\right)\gamma.
\end{equation}
Using the convexity of relative entropy, we have
\begin{equation}
D[p(t)\|\gamma]\leqslant e^{-\mu t}D[p(0)\|\gamma]
\end{equation}

For the later use, we also derive a lower bound on $D[p(t)\|\gamma]$.
We check the time derivative of $D\left[p\|\gamma\right]$
\begin{align}
\frac{d}{dt}D\left[p\|\gamma\right] & =\frac{d}{dt}\sum_{i}p_{i}\left(\ln p_{i}-\ln\gamma_{i}\right)\nonumber\\
 & =\sum_{i}\frac{d}{dt}p_{i}\left(\ln p_{i}-\ln\gamma_{i}\right)\nonumber\\
 & =\mu\sum_{i}\left(\mu_{i}-p_{i}\right)\ln\frac{p_{i}}{\gamma_{i}}\nonumber\\
 & =-\mu\left(D\left[p\|\gamma\right]+D\left[\gamma\|p\right]\right)\nonumber\\
& \geqslant -2\mu D\left[p\|\gamma\right],
\end{align}
where we have used $D\left[p\|\gamma\right]\geqslant D\left[\gamma\|p\right]$.
Because relative entropy is always non-negative, the integration yields
\begin{equation}
\label{eq:app-relentlower}
D[p(t)\|\gamma]\geqslant e^{-2\mu t}D[p(0)\|\gamma].
\end{equation}

\subsection{\label{App:secC2}%
Upper bound of relative entropy}
Let
\begin{equation}
q = e^{-\mu \tau /N}.
\end{equation}
The $D(\tau)$ in the constant-shifting protocol can be upper bounded by
\begin{align}
D(\tau)&=D\left[P^{N}\|\gamma^{N}\right] \nonumber\\
& \leqslant qD\left[P^{N-1}\|\gamma^{N}\right]\nonumber\\
 & \leqslant q^{2}D\left[P^{N-2}\|\gamma^{N}\right]+q\left(1-q\right)D\left[\gamma^{N-1}\|\gamma^{N}\right]\nonumber\\
 & \leqslant\dots\nonumber\\
 & \leqslant q^{N}D\left[P^{0}\|\gamma^{N}\right]+\sum_{k=1}^{N-1}q^{k}\left(1-q\right)D\left[\gamma^{N-k}\|\gamma^{N}\right]\nonumber\\
 & \leqslant q^{N}D\left[P^{0}\|\gamma^{N}\right]+\sum_{k=1}^{N-1}q^{k}\left(1-q\right)D\left[\gamma^{0}\|\gamma^{N}\right]\nonumber\\
 & =\left[q^{N}+\sum_{k=1}^{N-1}q^{k}\left(1-q\right)\right]D\left[\gamma^{0}\|\gamma^{N}\right]\nonumber\\
 & =qD\left[\gamma^{0}\|\gamma^{N}\right],
\end{align}
where we have used $D\left[P^{k-1}\|\gamma^{k}\right]\geqslant D\left[P^{k-1}\|\gamma^{k-1}\right]$,  $D\left[\gamma^{k}\|\gamma^{N}\right]\leqslant D\left[\gamma^{0}\|\gamma^{N}\right]$,
and $P^{0}=\gamma^{0}$. 
Therefore, $D(\tau)$ has an exponential scaling
\begin{equation}
\label{eq:rel-ent-up}
D\left[P^{N}\|\gamma^{N}\right]\leqslant e^{-\mu\frac{\tau}{N}}D\left[\gamma^{0}\|\gamma^{N}\right],
\end{equation}
when $N$ is fixed.

\subsection{\label{App:secC3}%
Exponential scaling of the relative entropy for fixed error}

The Eq.\eqref{eq:rel-ent-up} shows that $D(\tau)$ of the constant-shifting protocol is exponentially decaying up to a factor $D[\gamma(0)\|\gamma(\tau)]$.
This factor is a constant for the case of fixed energy.
Here, we also show that for the case of fixed error, although $\gamma(\tau)$ is changing for different $\tau$, still $D(\tau)$ decays exponentially.

Consider a protocol where the reset error is fixed.
This means that for different times the final energy of the state is dependent on the total time $\tau$ of the protocol.
Let $E_{\max}(\tau_1)$ be the final energy for the protocol with length $\tau$ and error $\epsilon$.
We have shown that for the same upper energy at longer total times the relative entropy term decays exponentially. 
As the protocol is longer, additionally, the final state will be nearer to the thermal state (that keeps fixed, as it only depends on the final energy).
The final error will thus be smaller than $\epsilon$ for that protocol.
This means that, if instead of the energy, we keep the final error fixed, we can drive the energy of the level to be set to zero slower, which will, at any time of the protocol with fixed error, drive the thermal state slower to the reset state, in comparison to the case of the protocol with fixed final energy.
It then directly follows that for the protocol with fixed error $\epsilon_1$ the final thermal state for any length of the protocol $\tau_2>\tau_1$ will be farther away of the reset state, than for the protocol with fixed maximal energy $E_{\max}$, while the final state will obviously be the same (as the error is fixed). 
This means that the final state will be nearer (by any distance measure) to the final thermal state.
Concluding, we find that the decay of the relative entropy term needs to be at least as fast in the case of protocols with fixed error as in the case of protocols with fixed final energy, where it is exponential; which yields the claim.

\subsection{\label{App:secC4}%
Exponential dacay of work penalty}

We check the total work cost
\begin{equation}
W=\sum_{k=0}^{N-1}P_1^{k}{\cal E}.
\end{equation}
For each time interval of thermalization $\Delta\tau=\tau/N$, the partial
swap model leads to
\begin{equation}
P_1^{k} =e^{-\mu\Delta\tau}P_1^{k-1}+\left(1-e^{-\mu\Delta\tau}\right)\gamma^{k}_1.
\end{equation}
The total work cost is then
\begin{align}
W & =\sum_{k=1}^{N}P_1^{k-1}{\cal E}\\
 & =\sum_{k=1}^{N}\gamma_1^{k-1}{\cal E}+\sum_{k=1}^{N}\left(P_1^{k-1}-\gamma_1^{k-1}\right){\cal E}\\
 & =\sum_{k=1}^{N}\gamma_1^{k-1}{\cal E}+e^{-\mu\Delta\tau}\sum_{k=2}^{N}\left(P_1^{k-2}-\gamma_1^{k-1}\right){\cal E}.\label{eq:app-wpn-discrete}
\end{align}

\begin{figure}[htbp]
\centering
\includegraphics[width=0.9\columnwidth]{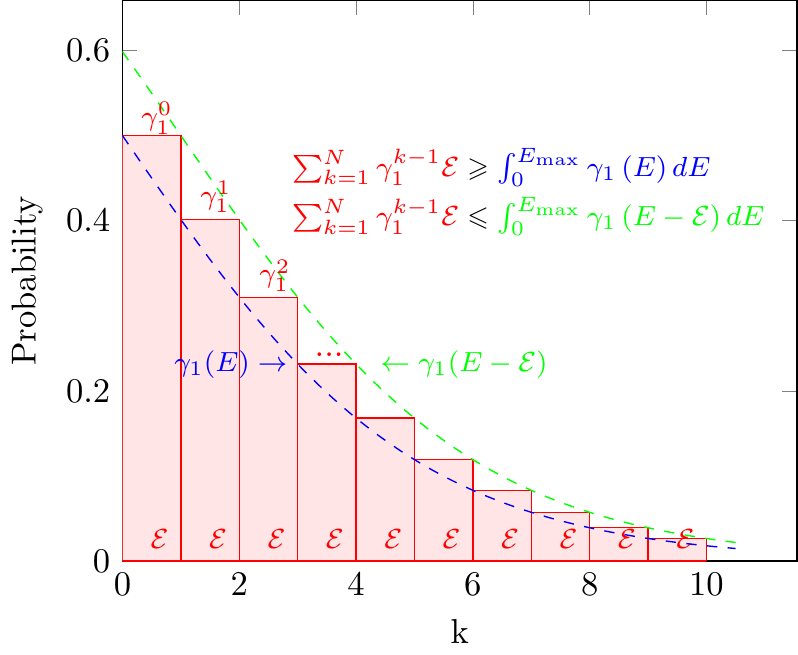}
\caption{\label{fig:app-sum2integ}Restrict sum by integral.}
\end{figure}

To calculate the exact value of $W$ is a difficult work. Instead, we
calculate the lower and upper bounds of $W$. The first term $\sum_{k=1}^{N}\gamma_1^{k-1}{\cal E}$
can be bounded by (see Fig.\ref{fig:app-sum2integ}
\begin{align}
\sum_{k=1}^{N}\gamma_1^{k-1}{\cal E}&\geqslant\int_{0}^{E_{\max}}\gamma_1\left(E\right)dE\label{eq:app-sum-int-l}\\
 \sum_{k=1}^{N}\gamma_1^{k-1}{\cal E}&\leqslant\int_{0}^{E_{\max}}\gamma_1\left(E-{\cal E}\right)dE.\label{eq:app-sum-int-u}
\end{align}
Substituting the above two equations into the work cost in Eq.\eqref{eq:app-wpn-discrete}, together with  $\gamma_1^k\leqslant P_1^{k}\leqslant 1/2$, we have bounds as
\begin{align}
W \geqslant & \int_{0}^{E_{\max}}\gamma_1\left(E\right)dE +e^{-\mu\Delta\tau}\sum_{k=2}^{N}\left(\gamma_1^{k-2}-\gamma_1^{k-1}\right){\cal E},\\
\geqslant & W_{\rm qs}+e^{-\mu\Delta\tau}\left(P_{0}^{{\rm eq}}-P_{N-1}^{{\rm eq}}\right){\cal E}\\
= & W_{\rm qs}+e^{-\mu\Delta\tau}\left(\frac{1}{2}-\frac{1}{1+e^{\beta\left(E_{\max}-{\cal E}\right)}}\right){\cal E},
\end{align}

\begin{widetext}
\begin{align}
W\leqslant & \int_{0}^{E_{\max}}\gamma_1\left(E-{\cal E}\right)dE+e^{-\mu\Delta\tau}\sum_{k=2}^{N}\left(\frac{1}{2}-\gamma_1^{k-1}\right){\cal E}=  \int_{-{\cal E}}^{E_{\max}-{\cal E}}\gamma_1\left(E\right)dE+e^{-\mu\Delta\tau}\left[\frac{N-1}{2}-\sum_{k=2}^{N}\gamma_1^{k-1}\right]{\cal E}\\
\leqslant & W_{\rm qs}+\left(\int_{-{\cal E}}^{0}-\int_{E_{\max}-{\cal E}}^{E_{\max}}\right)\gamma_1\left(E\right)dE+e^{-\mu\Delta\tau}\left[\frac{E_{\max}-{\cal E}}{2}-\int_{{\cal E}}^{E_{\max}}\gamma_1\left(E\right)dE\right]\\
= & W_{\rm qs}+T\ln\left[\frac{1}{2}+\frac{1}{2}\frac{e^{\beta{\cal E}}+e^{-\beta E_{\max}}}{1+e^{-\beta\left(E_{\max}-{\cal E}\right)}}\right]+e^{-\mu\Delta\tau}\left[\frac{E_{\max}-{\cal E}}{2}-W_{\rm qs}+\int_{0}^{{\cal E}}\gamma_1dE\right]\\
= & W_{\rm qs}+e^{-\mu\Delta\tau}\left[\frac{E_{\max}}{2}-W_{\rm qs}\right]+T\ln\left[\frac{1}{2}+\frac{1}{2}\frac{e^{\beta{\cal E}}+e^{-\beta E_{\max}}}{1+e^{-\beta\left(E_{\max}-{\cal E}\right)}}\right]+e^{-\mu\Delta\tau}\left(T\ln\frac{2}{1+e^{-\beta{\cal E}}}-\frac{{\cal E}}{2}\right).
\end{align}
\end{widetext}
The last term is always less than $0$. For the second term, let $x=e^{\beta{\cal E}}\geqslant1$
and $y=e^{-\beta E_{\max}}=x^{-N}$, using $\ln x\leqslant x-1$,
\begin{align}
& \ln\left[\frac{1}{2}+\frac{1}{2}\frac{e^{\beta{\cal E}}+e^{-\beta E_{\max}}}{1+e^{-\beta\left(E_{\max}-{\cal E}\right)}}\right]\nonumber\\ 
\leqslant & \frac{1}{2}\frac{x+y}{1+xy}-\frac{1}{2}=\frac{\left(x-1\right)\left(x^{N}-1\right)}{2\left(x^{N}+x\right)}\\
 & \leqslant\frac{x-1}{2}.
\end{align}
Finally, the work penalty can be characterized by $W_{\rm pn}^L\leqslant W_{\rm pn}\leqslant W_{\rm pn}^U$ with
\begin{align}
W_{\rm pn}^L & =e^{-\mu\tau/N}\left[\frac{1}{2}-\frac{1}{1+e^{\beta\left(E_{\max}-{\cal E}\right)}}\right]{\cal E}\label{eq:app-wpn-discrete-L}\\
W_{\rm pn}^U & =T\frac{e^{\beta{\cal E}}-1}{2}+e^{-\mu\tau/N}\left[\frac{E_{\max}}{2}-W_{\rm qs}\right].\label{eq:app-wpn-discrete-U}
\end{align}
Therefore, the work penalty decays exponentially to a non-zero constant.

\section{\label{App:secD}%
The partial swap rate of overdamped Fokker-Plank equation}

The dynamics of a Brownian particle in a double-well potential can be described by the Fokker-Planck equation
\begin{equation}\label{eq:app-FokkerPlanck}
\partial_tp\left(x,t\right) = D \partial_x\left[\beta\partial_x V\left(x,t\right)+\partial_x\right]p\left(x,t\right).
\end{equation}
where $p(x,t)$ is the distribution of the Brownian particle position at time $t$, $D$ is the diffusion constant, and $V(x,t)$ is the double-well potential.
Compared with the discrete case, the position $x$ casts as the continuous states label and the potential $V(x,t)$ the internal energies.
Therefore, by writing $\sum_i p_i f_i$ as $\int_{-\infty}^{+\infty}dx p(x)f(x)$, where $f$ can be internal energies (potentials) or stochastic entropies, we generalize the definitions in the discrete case to the continuous.
For instance, for $V(x,t)$ at time $t$, the corresponding thermal distribution is
\begin{equation}
\gamma\left(x,t\right)=\frac{e^{-\beta V\left(x,t\right)}}{Z},\quad Z\left(t\right)=\int_{-\infty}^{+\infty}dx e^{-\beta V(x,t)}.
\end{equation}
Then, it can be verified that the work penalty relation
\begin{align}
W_{\rm pn}\left(\tau\right) 
&= W\left(\tau\right) - W_{\rm pn}\\
&= T\Delta D\left[p\|\gamma\right]+T\Sigma\left(\tau\right),
\end{align}
naturally holds.

To perform the coarse-graining analysis, it is necessary to write the Fokker-Planck equation in the form of continuous master equation
\begin{equation}
\partial_t p\left(x,t\right) = \int_{\mathbb{R}} dx^{\prime} \Gamma\left(x|x^{\prime};t\right)p\left(x^{\prime},t\right).
\label{eq:app-mastereqcontin}
\end{equation}
Here, $\Gamma(x|x^{\prime};t)$ is the transition rate from $x^{\prime}$ to $x$.
It satisfies $\Gamma(x|x^{\prime};t)\geqslant 0$ $\forall x\neq x^{\prime},t$ and $\int dx \Gamma(x|x^{\prime};t)=0$ $\forall t$.
To write the differential form of Eq.\eqref{eq:app-FokkerPlanck} to the integral form of Eq.\eqref{eq:app-mastereqcontin}, we consider a small time interval
\begin{equation}
\delta t = t-t^{\prime}.
\end{equation}
Because the evolution is Markovian, the distribution $p(x,t)$ can be viewed as a one-dimensional random walk from $p(x,t^{\prime})$,
\begin{equation}
p\left(x,t\right) = \int_{\mathbb{R}}dx^{\prime} P\left[x,t|x^{\prime},t^{\prime}\right]p\left(x^{\prime},t^{\prime}\right),
\end{equation}
where $P[x,t|x^{\prime},t^{\prime}]$ is the jump probability from position $x^{\prime}$ at time $t^{\prime}$ to position $x$ at time $t$.
Then, we can approximate $\partial_t p(x,t)$ via the left derivative
\begin{equation}
\frac{p\left(x,t\right)-p\left(x,t^{\prime}\right)}{\delta t} = \frac{1}{\delta t}\int_{\mathbb{R}}dx^{\prime}\left(P\left[x,t|x^{\prime},t^{\prime}\right]-\delta_{xx^{\prime}}\right)p\left(x^{\prime},t^{\prime}\right).
\end{equation}
Compared with Eq.\eqref{eq:app-mastereqcontin}, we can approximate $\Gamma(x|x^{\prime};t)$ by
\begin{equation}
\Gamma\left(x|x^{\prime};t\right) = \lim_{\delta t \rightarrow 0}\frac{P\left[x,t|x^{\prime},t^{\prime}\right]-\delta_{xx^{\prime}}}{\delta t}.
\label{eq:app-overdampedGamma}
\end{equation}
Here we assume that the above limitation is well-defined.
It this is not the case, we could always use $P[x,t|x^{\prime},t^{\prime}]$ of small time $\delta t$ to evaluate $\Gamma(x|x^{\prime};t)$.

From the Fokker-Planck equation in Eq.\eqref{eq:app-FokkerPlanck}, $P[x,t|x^{\prime},t^{\prime}]$ can be solved as~\cite{Seifert12}
\begin{equation}
P\left[x,t|x^{\prime},t^{\prime}\right] = 
\frac{1}{\sqrt{4\pi D\delta t}}
e^{-\frac{\left(\delta x+\beta D \partial_{x^{\prime}}V\left(x^{\prime},t^{\prime}\right) \delta t\right)^2}{4D\delta t}}.
\end{equation}
Upon Eq.\eqref{eq:app-overdampedGamma}, we can verify that the detailed balanced condition holds, i.e.
\begin{equation}
\Gamma\left(x|x^{\prime};t\right)\gamma\left(x^{\prime},t\right) = \Gamma\left(x^{\prime}|x;t\right)\gamma\left(x,t\right).
\end{equation}
This allows us to apply the coarse-grained entropy production in Eq.\eqref{eq:app-entprodcoarsegrain} as a lower bound of the actual entropy production, as shown in Eq.\eqref{eq:app-entprod-entprodcg}.
Here, the coarse-grained states are
\begin{align}
P_0\left(t\right) &= \int_{-\infty}^0 dx p\left(x,t\right),\\
P_1\left(t\right) &= \int^{+\infty}_0 dx p\left(x,t\right).
\end{align}
That is, when using the Brownian particle in double-well potential as a bit, we choose position in the left well as bit value 0 while position in the right well as bit value 1.

Analogy to the discrete case, the coarse-grained bit system satisfies mater equation with transition rate
\begin{align}
\Gamma_{01}^{\rm bit} &=\int_{-\infty}^0 du \int_0^{+\infty}dv\Gamma\left(u|v;t\right)\frac{p\left(v,t\right)}{P_1\left(t\right)},\\
\Gamma_{10}^{\rm bit} &=\int_0^{+\infty}dv \int_{-\infty}^0du \Gamma\left(v|u;t\right)\frac{p\left(u,t\right)}{P_0\left(t\right)}.
\end{align}
The dynamics can also be described by the partial swap model as in Eq.\eqref{eq:app-partialswap}.
Particularly, the partial swap rate $\mu=\Gamma_{01}^{\rm bit}+\Gamma_{10}^{\rm bit}$ can be well estimated.

\section{\label{App:secE}%
Implied bound on information throughput}
As a two level system optimizes the work penalty, here we consider the bit is constituted by truly two-level systems for the information throughput.
By substituting
\begin{align}
&\ln\frac{Z(0)}{Z(\tau)} +D(\tau)\\
=& \ln2- \ln(1+e^{-\beta E_{\max}}) + \epsilon\ln\frac{\epsilon}{1/(1+e^{\beta E_{\max})}}\\
&+ (1-\epsilon)\ln\frac{1-\epsilon}{1/(1+e^{-\beta E_{\max}})}\\
=&\ln2 - H_b(\epsilon)+ \epsilon\ln\frac{1+e^{\beta E_{\max}}}{1+e^{-\beta E_{\max}}}\\
=&\ln2 - H_b(\epsilon)+ \epsilon\beta E_{\max},
\end{align}
where $H_b(\epsilon)=\epsilon\ln\epsilon+(1-\epsilon)\ln(1-\epsilon)$ is the binary entropy, into the binary switch energy
\begin{align}
E_{\rm bit}&=W_{\rm qs}+W_{\rm pn}\\
&=\kb T\ln\frac{Z(0)}{Z(\tau_{\rm SW})} +\kb TD(\tau_{\rm SW}) +T\Sigma(\tau_{\rm SW}),
\end{align}
we obtain
\begin{align}
P&=\frac{E_{\rm bit}}{B}\\
&\geqslant \frac{\kb Tn}{\tau_{\rm SW}}\left(\ln2 - H_b(\epsilon)+ \epsilon\beta E_{\max}+\frac{(1-2\epsilon)^2}{\mu \tau_{\rm SW} }\right).
\end{align}

\bibliography{ftlrefs}